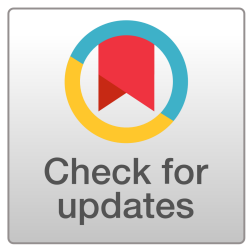
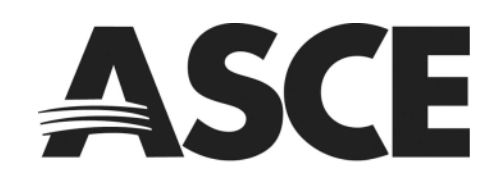

# A Cost-Effective Slag-Based Mix Activated with Soda Ash and Hydrated Lime: A Pilot Study

Jayashree Sengupta, S.M.ASCE[1]; Nirjhar Dhang[2]; and Arghya Deb[3]

**Abstract:** The present study explores a cost-effective method for using activated ground granulated blast furnace slag (GGBFS) and silica fume (SF) as cement substitutes. Instead of activating them with expensive alkali solutions, the present study employs industrial-grade powdered sodium aluminate (SA) and hydrated lime (HL) as activators, reducing expenses by about 94.5% compared to their corresponding analytical-grade counterparts. Herein, the exclusivity is depicted using less pure chemicals rather than relying on reagents with 99% purity. Two mixing techniques are compared: one involves directly introducing powdered SA and HL, while the other premixes SA with water before adding it to a dry powder mixture of GGBFS, SF, and HL. Microstructural analysis reveals that the initial strength results from various hydrate phases, including calcium–sodium–aluminate–silicate hydrate. The latter strength is attributed to the coexistence of calcium–silicate hydrate, calcium–aluminate–silicate hydrate, and sodium–aluminate–silicate hydrate, with contributions from calcite and hydrotalcite. The SF content significantly influenced the formation of these gel phases. Thermogravimetric analysis (TGA) reveals phase transitions and bound water related to hydration products. The optimal mix comprises 10% SF, 90% GGBFS, 9.26% HL, and 13.25% SA, with a water-to-solid ratio of 0.45. This approach yields a compressive strength of 35.1 MPa after 28 days and 41.33 MPa after 120 days, hence being suitable for structural construction. **DOI: 10.1061/PPSCFX.SCENG-1426.** *© 2024 American Society of Civil Engineers.*

## Introduction

Natural aluminosilicates have been combined with limestones since time immemorial, creating durable building materials, with Roman concrete serving as a notable example (Jackson et al. 2017; Seymour et al. 2022). On similar grounds, the recent technological advancement of alkali activation of industrial byproducts, such as, ground granulated blast furnace slag (GGBFS), aims to produce "green" binders, often referred to as "alkali-activated binders" (Provis 2018; Shi et al. 2011). These binders yield concrete similar to traditional cement but possess different microstructural compositions depending on the raw materials used. Consequently, alkali activation valorizes industrial and agricultural waste materials, potentially reducing or even eliminating the need for conventional cement. Such a transition from cement can efficiently reduce airborne carbon dioxide emissions (Flower and Sanjayan 2007; Teh et al. 2017; Wu et al. 2018).

The core concept behind alkali activation involves the transformation of nonreactive aluminosilicates, termed precursors, into binding agents when exposed to alkalis. These alkalis are typically sodium or potassium-based and are often used in combination with silicate-based solutions (Jagad et al. 2023a, b). Alkaline earth metal-based alkalis, such as calcium and magnesium, have the potential to enhance mechanical strength and durability by forming additional calcium-rich hydrates (Yip et al. 2005).

The working mechanism follows that the aluminosilicate precursors disintegrate with the addition of alkalis, releasing tetrahedral monomers. The excess oxygen is then shared between the metal ions (sodium, calcium, or both) and alumina, forming a polymeric gel with stronger bonds, thus enhancing the strength. Divalent calcium ions can react with hydroxyl groups to form precipitates, elevating the pH and creating sites for the nucleation and polymerization of soluble silicates (Lee and Van Deventer 2002). It is important to note that the continuous formation of binding gels such as calcium–silicate hydrate (CSH), calcium–aluminate–silicate hydrate (CASH), sodium–aluminate–silicate hydrate (NASH), and calcium–sodium–aluminate–silicate hydrate (CNASH) relies heavily on the availability of calcium and sodium ions as well as the pH level within the mixture.

Despite the advantages, alkali activation has not yet been adopted into standard concrete practice, primarily due to (1) the need for trained supervision when using alkalis in solution form (Adesanya et al. 2018; Nematollahi et al. 2015); (2) the fact that alkali solutions are highly skin-corrosive and can result in first-degree burns if mishandled (IPCS 1998); (3) the use of silicates increases $CO_2$ emissions (Fawer et al. 1999); (4) the cost of the alkalis (Table 2); and (5) the corresponding storage requirements, which adds to the budgetary burden of construction.

An alternative approach to address these challenges is the utilization of alkalis in solid form, referred to as "one-part" alkali-activated concreting (Luukkonen et al. 2018). Unlike the traditional two-part mixing involving alkali solutions, in a one-part system, a solid alumino-silicate precursor is combined with a powdered alkali activator, followed by the addition of water. This approach aligns more closely with customary concrete mixing practices. In one-part mixing, the primary reaction centers around ion exchange during hydrolysis, which precedes the dissolution and release of silica and alumina monomers, ultimately leading to polymerization similar to that observed in the two-part mixing process.

[1]Research Scholar, Dept. of Civil Engineering, Indian Institute of Technology, Kharagpur, West Bengal 721302, India (corresponding author). ORCID: https://orcid.org/0000-0002-6892-3480. Email: jaish.sengupta@iitkgp.ac.in

[2]Professor, Dept. of Civil Engineering, Indian Institute of Technology, Kharagpur, West Bengal 721302, India.

[3]Professor, Dept. of Civil Engineering, Indian Institute of Technology, Kharagpur, West Bengal 721302, India.

Two key factors that significantly influence the properties of alkali-activated mixes, both in their fresh and hardened states, are the choice of aluminosilicate precursor and the activator. It is important to note that the availability of industrial byproducts varies by location, which can present constraints. The dissolution characteristics of the precursors are contingent upon their type, fineness, and mineral composition. Mechanochemical processing of aluminosilicates can accelerate early strength development, as observed in the study by Kearsley et al. (2015). However, such preprocessing steps may impact cost-effectiveness. In addition to the choice of precursors, the concentration of the activator and the mixing procedure also play crucial roles in determining the mechanical properties of alkali-activated concrete. Alkali metal hydroxides, specifically sodium and potassium hydroxides ($Na^+$, $K^+$), are commonly used as activators due to their ability to promote dissolution at a higher pH level within the system, as highlighted by Hardjito and Rangan (2005). Nonetheless, it is worth noting that even in solid form, these hydroxides are highly caustic. Furthermore, they are typically used in conjunction with sodium silicate, which can be expensive, necessitating the exploration of milder and more cost-effective alternative activators.

Many researchers have explored alternative activators made from silica-rich wastes, such as rice husk ash (RHA), ground waste glass (WG), and silica fume (SF), to replace sodium silicate solutions. These activators are designed by dissolving waste products in NaOH or KOH-concentrated solutions. The RHA-based activator showed similar strength development as sodium silicate (Handayani et al. 2022), but higher RHA content may affect themix's flow characteristics (Athira and Bahurudeen 2022). The WG-NaOH-activated mix had higher compressive strength than a silicate-activated mix. Silica fume was more effective than RHA (Vigneshwari et al. 2018), as it benefited from its filling action, creating more calcium silicate hydrates (Kang et al. 2019; Nežerka et al. 2019; Rostami and Behfarnia 2017), increasing strength and durability. However, higher amounts of SF reduced strength (Cheah et al. 2019) and increased the volume of large capillary pores (Li et al. 2022). The optimal SF amount was found to be 10%. Incorporating sugarcane straw ash into NaOH and preconditioning it for 24 h yielded strengths exceeding 50 MPa, as demonstrated by Moraes et al. (2017). Pastes consisting of olive biomass ash and GGBFS exhibited a 28-day mechanical strength ranging from 18 to 33 MPa (Alonso et al. 2019). However, it is important to note that caustic alkalis were still utilized in the process.

Using analytical-grade $Na_2CO_3$ has yielded higher later-age strength with a denser microstructure (Abdalqader et al. 2016) and excellent acid resistance (Abdalqader et al. 2019). A $Na_2CO_3$-activated GGBFS binder produced C─S─H with riversiderite structure and $CaCO_3$. The interaction of $Na^+$, $CO_3^{2-}$, and $Ca^{2+}$ ions resulted in the formation of gaylussite [$Na_2Ca(CO_3)2H_2O$]. Also, being less caustic, handling $Na_2CO_3$ makes it more suitable for on-site construction. However, $Na_2CO_3$ alone may not sufficiently elevate the system's pH to sustain the dissolution mechanisms, necessitating supplementation with another alkali, preferably milder, to maintain a higher pH threshold. This supplementation can be achieved by using calcium hydroxide simultaneously with sodium carbonate, leading to the sequential formation of sodium hydroxide and calcium carbonate, as proposed by Adesina (2020). Pore refinement was reported by adding $Na_2CO_3$ to the $Ca(OH)_2$-activated fly ash mix (Jeon et al. 2015). Akturk et al. (2019) documented a 28-day strength development ranging between 28 and 45 MPa for different $Na_2CO_3$─$Ca(OH)_2$─NaOH combinations.

Prior research has demonstrated that utilizing analytical-grade chemicals can enhance the mix properties. Still, there has been a lack of investigation into the possible benefits of less pure industrial-grade chemicals such as soda ash and hydrated lime. These industrial-grade activators offer cost advantages due to their lower purity. Therefore, there is potential to explore their use as substitutes and assess their impact on strength development in structural applications. In this context, the current study focuses on the utilization of soda ash (SA) and hydrated lime (HL). Soda ash, which is industrial-grade sodium carbonate, finds extensive use in various industries, including glass, detergent, paper, and chemical manufacturing. It is also used in water treatment and as a food additive, making it widely accessible globally. Industrial-grade hydrated lime is employed in water and waste treatment as well as construction processes, with varying purity levels ranging from 75% to 96% depending on the CaO content.

Consequently, the primary objective of the present study is to establish the applicability of industrial-grade chemicals in alkali activation. The study explores and advocates for a one-part mixing methodology involving the direct combination of industrial SA and HL with a mixture of GGBFS and silica fume (SF). The study assesses the impact of this approach on both the fresh and hardened properties of the mixtures and conducts microstructural characterization of selected mortar samples to validate the results obtained from the one-part mixing process. Finally, the study summarizes its significant contributions. It is anticipated that this research would make valuable contributions to ongoing efforts aimed at developing sustainable and economically viable solutions for construction materials, thereby advancing the field of alkali-activated binders.

## Research Significance

Two different mixing techniques are employed: one involves introducing powdered forms of SA and HL [Fig. 3(a)], and the other where SA is premixed with water before being added to a dry powdered mixture of GGBFS, SF, and HL [Fig. 3(c)]. While a previous study (Kearsley et al. 2015) reported optimal strength with a 6% designed NaOH solution, this study does not yield the same results, suggesting that mixing techniques and activator quality, precisely purity, significantly influence the properties of the resulting mixes. Furthermore, this research showcases a substantial cost difference when using industrial-grade chemicals compared to analytical-grade $Na_2CO_3$ and $Ca(OH)_2$, potentially saving up to 94.5% of the budget, as tabulated in Tables 1 and 2. A cost saving of 91% is seen when contrasting the SA–HL activated mixture with a stoichiometrically equivalent two-part NaOH solution-activated mixture.

This study aims to address the challenges associated with *in situ* casting using alkali-activated binders and offer a plausible solution. While previous studies often focus on expensive or hazardous chemicals, this study emphasizes the need for an activator that balances mechanical strength with user-friendliness, accessibility, and affordability. The novelty lies in the use of less pure chemicals (80%–90% purity) added in solid form rather than relying on reagents with 99% purity. Fig. 1 portrays the research significance, highlighting this research's practical and cost-effective methodology.

**Table 1.** Cost comparisons for analytical- and industrial-grade chemicals for 9.26% of HL and 13.25% of SA, corresponding to a total of 10% NaOH for 1 $m^3$ of mortar mix in the Indian context

| Industrial-grade soda ash (as solid or premixed in water) and hydrated lime (as solid) ($INR/m^3$) | Analytical-grade NaOH pellets in water ($INR/m^3$) | Analytical-grade $Na_2CO_3$ and $Ca(OH)_2$ added in powder form ($INR/m^3$) |
|---|---|---|
| 3,428.49 | 39,426.60 | 63,115.13 |

**Table 2.** Costing schedule of the analytical- and industrial-grade chemicals as available in the market, in the Indian context

| Types of chemicals | Measured as | Cost | Ref. |
|---|---|---|---|
| Analytical-grade NaOH | INR per kg | 690.00 | Merck (2023c) |
| Analytical-grade $Na_2CO_3$ | INR per kg | 610.00 | Merck (2023b) |
| Analytical-grade $Ca(OH)_2$ | INR per kg | 320.00 | Merck (2023a) |
| Industrial-grade soda ash | INR per kg | 35.50 | Tata Chemicals Limited (2022) |
| Industrial-grade hydrated lime | INR per kg | 14.00 | Shreeram Chemical Industries (2022) |

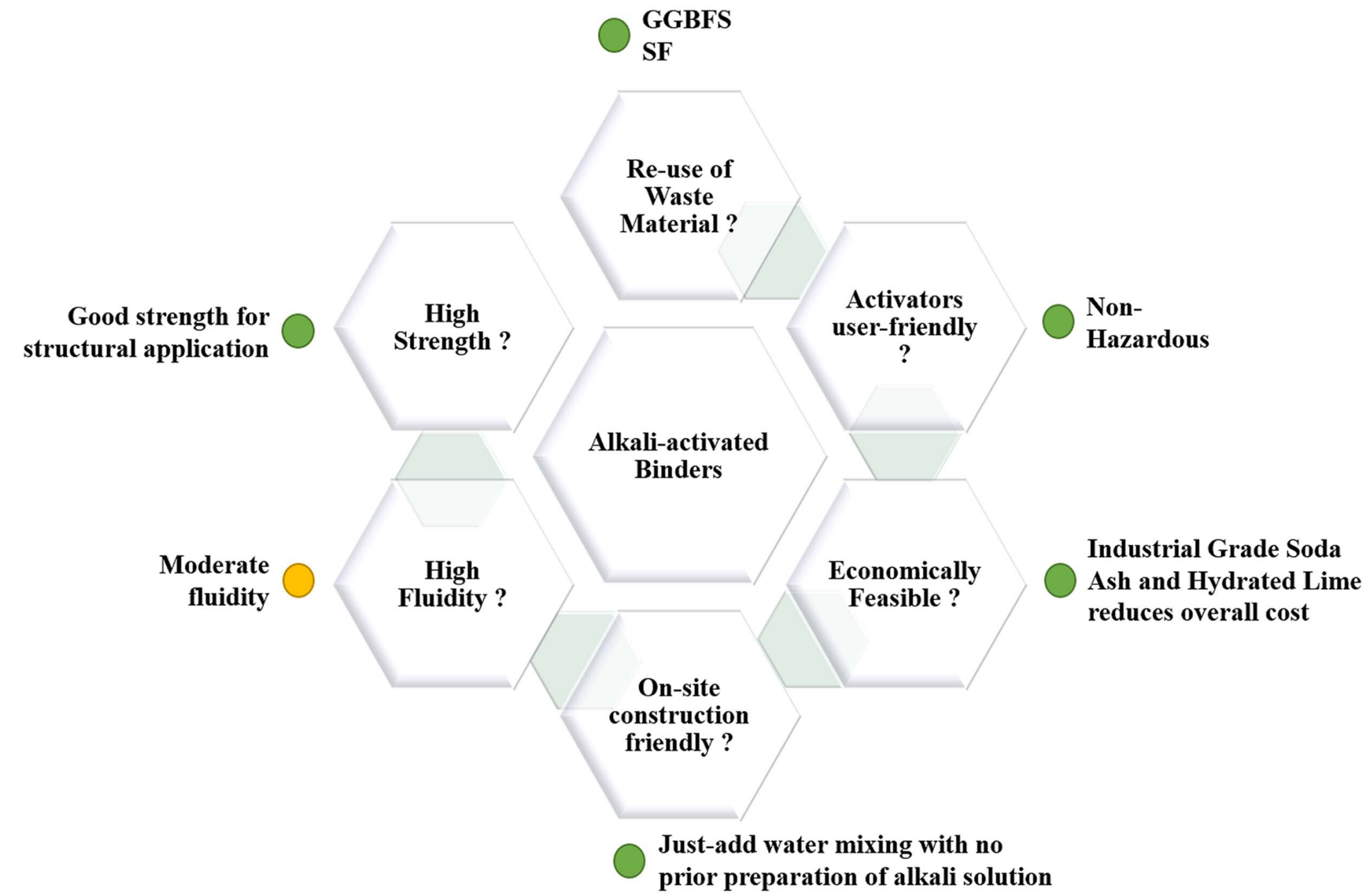

**Fig. 1.** Identifying the importance of the research and the likelihood of its implementation in practice.

## Material and Experimental Methods

### *Material Characterization*

GGBFS from Rashmi Cement Limited, West Bengal, India, was used as the aluminosilicate precursor. Silica fume (SF) was obtained from Waltar Enterprises. Table 3 lists the chemical compositions as detected by x-ray fluorescence spectroscopy analysis. Shreeram Chemicals Kolkata supplied the industrial-grade powdered hydrated lime (HL, LOI is 25.70%) and soda ash (SA, LOI is 8.46%). The mineralogical compositions of the activators were determined from x-ray diffraction (XRD) by Rietveld refinement analysis on X'Pert Highscore Plus 3.0 software, signifying the lower purity of the chemicals, and are shown separately in Fig. 2.

The reactivity of steel slag depends on the modulus of basicity (B). The modulus of basicity of GGBFS is the ratio between the total basic oxides and the total acidic oxides. The slag can be categorized as acidic ($B < 1$), neutral ($B = 1$), and basic ($B > 1$). For alkali activation, the slag has to be either neutral or basic. Acidic slags are better suited for pozzolanic application. The acidic oxides ($O_2^-$ ions acceptors) in GGBFS are $SiO_2$ and $P_2O_5$, whereas the alkali metal ($K_2O$ and $Na_2O$) and alkali–earth metal (CaO, MgO, and $Fe_2O_3$) oxides represent basic oxide ($O_2^-$ ions donators) and $Al_2O_3$, an amphoteric oxide. Amphoteric oxides act as acidic oxides in the presence of strong bases. With the $Fe_2O_3$, $K_2O$, $Na_2O$, and $P_2O_5$ present in a negligible amount, B equals $(CaO + MgO)/(SiO_2 + Al_2O_3)$. Also, the GGBFS requires a criterion of a $CaO/SiO_2$ ratio < 1.4 [as per BS: 6,699 (BSI 1992)] to act as a precursor material. The slag experimented on in this paper has $B = 1.14$ and $CaO/SiO_2 = 1.36$, satisfying the criteria to be used as a precursor for alkali activation. Standard sand conforming to IS 650:1991 (ISO 1991) was used as the fine aggregate throughout. The NaOH, wherever used, was prepared from 98% pure pellets procured locally.

### *Mix Proportions*

Table 4 lists the mix proportions of the mortar samples cast. The SF substituted the GGBFS for all sets of binders. The first set, set A, showed both activators being added in powder form, resulting in one-part alkali activation. The powdered SA and HL were mixed

**Table 3.** Properties of the precursor and additive

| Binders | CaO | $SiO_2$ | $Al_2O_3$ | MgO | MnO | $K_2O$ | $Na_2O$ | $Fe_2O_3$ | $TiO_2$ | $P_2O_5$ | $SO_3$ | LOI | Density ($kg/m^3$) |
|---|---|---|---|---|---|---|---|---|---|---|---|---|---|
| GGBFS | 43.78 | 32.08 | 11.20 | 5.82 | 0.84 | 0.42 | 0.03 | 0.75 | 0.87 | 1.33 | 1.65 | 2.74 | 2,890 |
| SF | 3.81 | 84.12 | 0.15 | 1.43 | 1.15 | 2.70 | 0.02 | 2.64 | 0.40 | 0.72 | 0.33 | 3.41 | 2,170 |

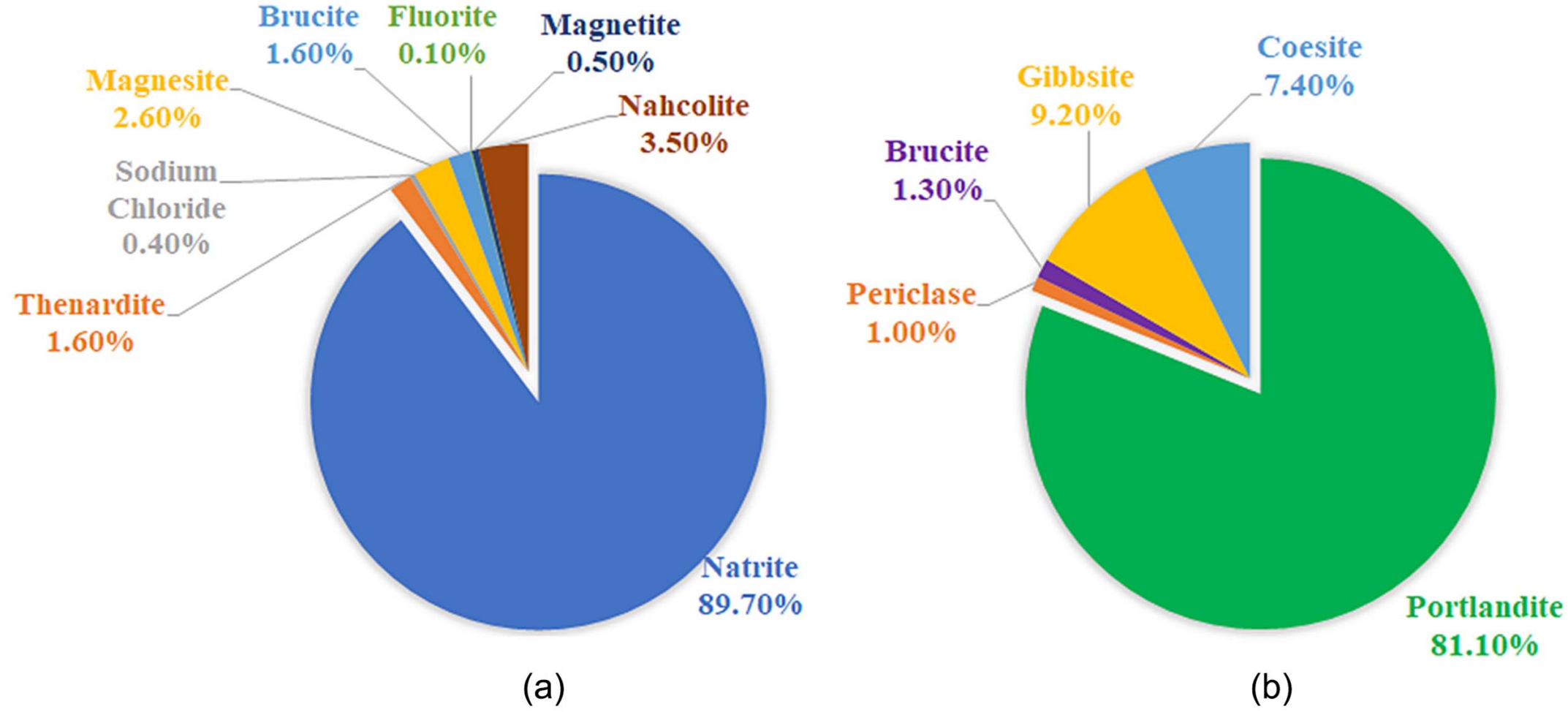

**Fig. 2.** Mineralogical compositions showing the impurities in (a) soda ash; and (b) hydrated lime.

**Table 4.** Mix proportion of mortars and the cost of binders/kg of binder (INR/kg)

| Set type and sample description | Mix ID | Binder | | Activator added per 100 g of binder | | | | Sand (g) | Cost of activators/kg of binder (INR/kg) | Remarks |
|---|---|---|---|---|---|---|---|---|---|---|
| | | SF (g) | GGBFS (g) | NaOH (g) | Targeted NaOH (g) | Hydrated lime (g) | Soda ash (g) | | | |
| Set-A: soda ash and hydrated lime added in solid form | SF10NH6 | 10 | 90 | — | 6 | 5.55 | 7.95 | 300 | 3.6 | See note |
| | SF20NH6 | 20 | 80 | | | | | 300 | | |
| | SF10NH8 | 10 | 90 | — | 8 | 7.41 | 10.59 | 300 | 4.8 | |
| | SF20NH8 | 20 | 80 | | | | | 300 | | |
| | SF10NH10 | 10 | 90 | — | 10 | 9.26 | 13.25 | 300 | 6 | |
| | SF20NH10 | 20 | 80 | | | | | 300 | | |
| | SF10NH12 | 10 | 90 | — | 12 | 11.11 | 15.9 | 300 | 7.2 | |
| | SF20NH12 | 20 | 80 | | | | | 300 | | |
| Set-B: control NaOH solution | SF10NH8_C | 10 | 90 | 8 | — | — | — | 300 | 55.2 | For a w/s = 0.45, NaOH is of 4M |
| | SF20NH8_C | 20 | 80 | | | | | 300 | | |
| | SF10NH10_C | 10 | 90 | 10 | — | — | — | 300 | 69 | For a w/s = 0.45, NaOH is of 5.2M |
| | SF20NH10_C | 20 | 80 | | | | | 300 | | |
| Set-C: pre-mixed soda ash in water | SF10NH8_PM | 10 | 90 | — | 8 | 7.41 | 10.59 | 300 | 4.8 | Same as Set-A mixes |
| | SF20NH8_PM | 20 | 80 | | | | | 300 | | |
| | SF10NH10_PM | 10 | 90 | — | 10 | 9.26 | 13.25 | 300 | 6 | |
| | SF20NH10_PM | 20 | 80 | | | | | 300 | | |

Note: $Na_2CO_3 + Ca(OH)_2 \rightarrow 2NaOH + CaCO_3$. The ratio of the molar mass of $Ca(OH)_2$ to the molar mass of 2NaOH is 74/80, and the ratio of the molar mass of $Na_2CO_3$ to the molar mass of 2NaOH is 106/80. SF10NH6 represents 10% of silica fume and 6% of equivalent NaOH resulting from the corresponding HL and SA reactions. SF10NH6_C represents 10% of silica fume and 6% of NaOH solution. SF10NH6_PM represents 10% of silica fume and 6% of premixed SA in water.

proportionately to achieve a stoichiometrically balanced amount of NaOH, as shown in Eq. (1). So, in order to form 10% NaOH, 9.26% of HL and 13.25% of SA were required to mix. The total solids summed up the aluminosilicate precursor, the additive, and the activators. The water-to-solid (w/s) ratio was based on the total solids ratio. Based on the trials, a minimum of w/s = 0.45 was required to meet the minimum water demand of the total mix. A w/s ratio of less than 0.45 resulted in a very dry mix, segregating the fine aggregates:

$$Na_2CO_3 + Ca(OH)_2 \rightarrow 2NaOH + CaCO_3 \qquad (1)$$

Additionally, samples were also cast for w/s of 0.5 and 0.55. Similarly, set B represented the two-part control mixtures (suffixed C), where the NaOH solution was added. Whereas set-C described, the SA pre-mixed in water (suffixed PM) was added to a mix of GGBFS-SF-HL combination. All the mortar samples were cast at room temperature.

## *Experimental Methods and Analytical Techniques*

### Paste Rheology

The study used an Anton Paar MCR 302 rheometer to conduct rheological tests on the corresponding paste samples for sets A, B, and C. The pastes were tested three times at a constant temperature of 32°C ± 0.5°C. The rheological shearing protocol involved initial preshearing at 100 $s^{-1}$ for 30 s to bring the paste to a steady state, followed by a 45-s rest period, and stepped ramps with different shear rates, varying from 0 to 100 to 0 $s^{-1}$, with each shear rate maintained for 20 s. The flow curve was determined from the up and down shear rate stepped ramps to assess the rheological properties. The modified Bingham (MB) model ($\tau = \tau_0 + \mu_p \dot{\gamma} + C\dot{\gamma}^2$) was employed to determine yield stress values, linking shear stress ($\tau$) to yield stress ($\tau_0$), plastic viscosity ($\mu_p$), regression constant (C), and shear rate ($\dot{\gamma}$). This model accommodated the nonlinear nature of the experimental rheological data, serving as an extension of the Bingham model. The advantage of this model lay in its ability to account for shear thinning, shear thickening, and Bingham fluids, depending on the $C/\mu_p$ ratio. Specifically, $C/\mu_p < 0$ indicated shear thinning, $C/\mu_p > 0$ implied shear thickening, and $C/\mu_p = 0$ signified Bingham fluids. Consequently, a negative regression constant C indicated shear thinning behavior, and only flow curves with a strong linear fit ($R^2 > 0.95$) were considered.

### Mortar Flow

Mortar specimens were put through flowability testing using the flow test. Selected samples were tested at various time intervals immediately after mixing and at 5, 10, 20, and 30 min postmixing. At the specified time points, a truncated cone-shaped mold measuring 70 mm × 100 mm × 50 mm was filled with the corresponding mortar samples and emptied. The sample was subjected to 25 jolts in 15 s on a flow table, and the average spread diameters of the mortars were measured in three different directions. These mortar compositions consisted of 400 g of GGBFS and SF, as detailed in Table 4, with varying water-to-solid (w/s) ratios of 0.45, 0.50, and 0.55. The flow percentage was calculated as $\text{Flow}(\%) = (\frac{D_t}{D_0} - 1) \times 100\%$, where, $D_t$ = obtained average spread diameter at time $t$; and $D_0$ = original base diameter of mold, i.e., 100 mm.

### Compressive Strength Test

The mortar samples were cast in 70.6 mm cubic moulds per IS 4031-6: 1988 (ISO 1988). A Hobart mixer was used to mix following the sequence, as shown in Figs. 3(a–c). The mortar was then placed in the mould on a table of a vibrating machine. The mortar was prodded 20 times in two layers to prevent honeycombing and voids. Forty-eight mixtures were cast in total, with nine replicas each. Samples were demolded after 24 h, cured at ambient temperature and humidity, and tested at 7, 28, and 120 days. Compressive strengths were determined using a Tinius Olsen Model 120 hydraulic Super ‘L’ universal testing machine with a 60-t load capacity.

### Sample Preparation for Phase Characterization

After the samples were tested for compressive strength, powdered mortar pieces were collected. The samples were dipped in acetone for 45 min and then air-dried for 5 min. The air-dried samples were then dried in an oven for 2 h at 60°C. The samples were then taken and kept in a vacuum desiccator until testing for XRD, scanning electron microscopy (SEM), and thermogravimetric analysis (TGA). Mixes with the highest compressive strength were analyzed for chemical composition using microstructural, XRD, and thermal analysis.

### Scanning Electron Microscopy–Energy Dispersive Spectrometry

Semiquantitative chemical composition was derived from the hardened sample intersections. Microstructural analysis was performed on gold-coated samples using a field emission-gun scanning electron microscope and energy dispersive spectrometer (SEM-EDS, ZEISS Merlin Scanning Electron Microscope with Oxford EDS Detector). EDS analyses were performed on 20 random points per sample, determining the molar ratios. The analysis was done using a backscatter electron detector with a 10 kV acceleration voltage.

### X-Ray Diffraction

Identification of crystalline phases for the hydrated products at 7 and 28 days was done with XRD using a Malvern Panalytical diffractometer (Cu anode material, K-alpha1/K-alpha2 = 0.50, K-beta $\lambda = 1.39225$ Å). The sample was scanned in the angular range of 8°–80° at a step of 0.02° (2$\theta$), and the time per step was 1 s. The profile fitting of the phases was performed in X'Pert HighScore Plus 3.0 software. The intensities calculated from a model of the crystalline structure are close-fitted to the observed x-ray pattern by the least-squares method, which is again obtained by varying the parameters of the crystal structure and the peak profiles to minimize the difference between the observed and the calculated patterns.

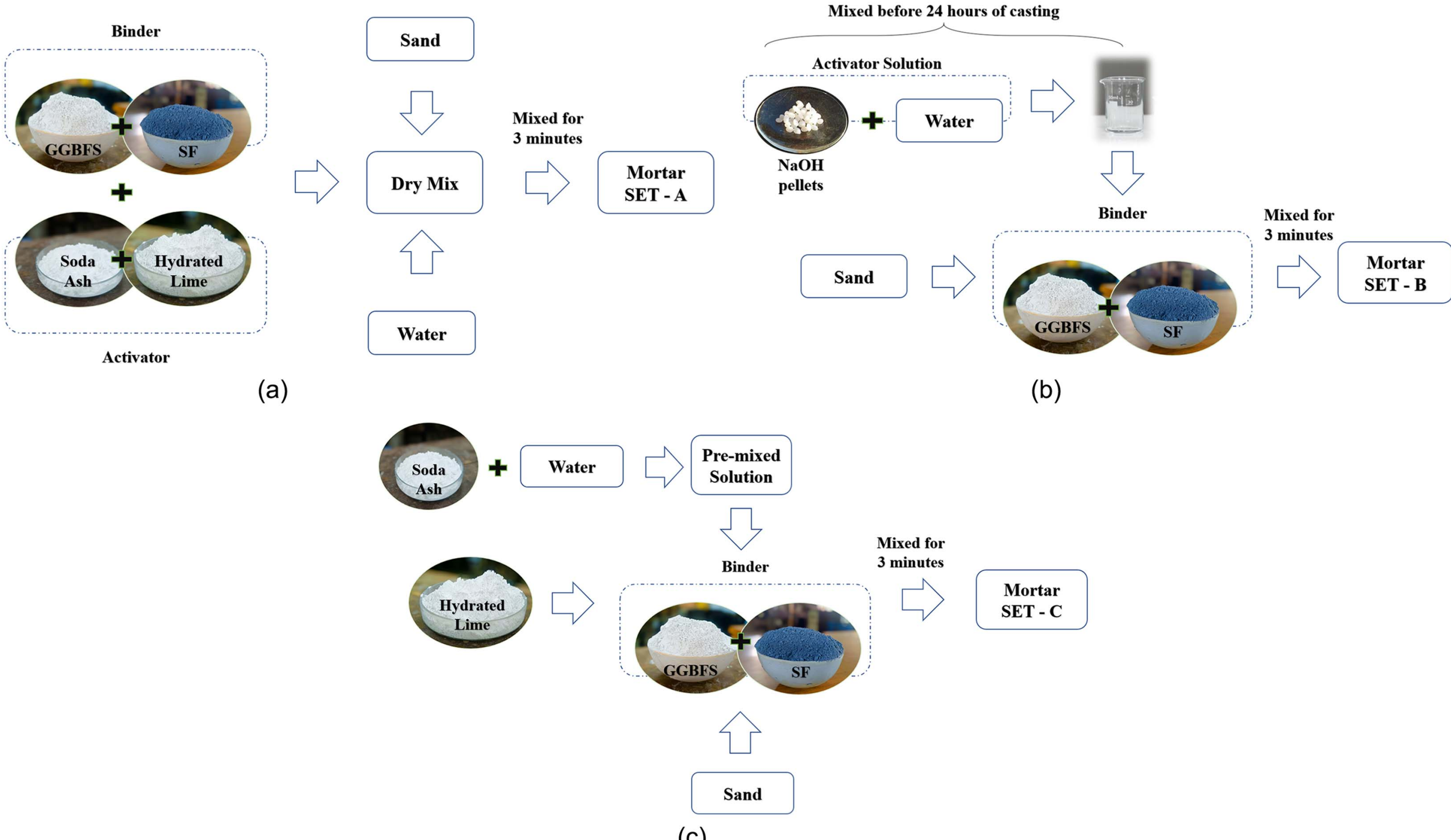

**Fig. 3.** Schematic diagram of the mixing sequence of the three different sets of mortar specimens (a) Set A; (b) Set B; and (c) Set C. (Images by authors.)

**Thermo-Gravimetric Analysis–Differential Thermal Analysis** Differential thermal analysis (DTA) and TGA were carried out simultaneously in a Pyris diamond thermal analyzer to measure the mass loss, which reflects the level of hydration processes, in powdered samples obtained from mortar cubes after conducting a compressive strength test. For the DTA-TGA analysis, 20 mg of the powdered sample with a particle size smaller than 45 mm was utilized. During the experimental setup, the mortar sample was subjected to gradual heating, starting from 27°C and increasing at 10°C per minute, all within a nitrogen ($N_2$) atmosphere. The chosen temperature range was designed to cover the region with minimal or absent mass loss. As the sample was heated, it underwent dissociation, releasing metal and hydroxyl groups. With further heating, chemically bound water would be expelled. Therefore, by observing the mass loss, precisely the TG value, we could estimate the progression of the reaction process. Mass loss was assessed using a TG balance with an accuracy of 0.1 mg.

## Results and Discussion

### Paste Rheology

Fig. 4 shows the yield stress and the plastic viscosity of the different pastes. Raising the targeted percentage of NaOH led to a decrease in both yield stress and plastic viscosity. However, once the additions surpassed 10%, the yield stress started to increase, as depicted in Fig. 4. Previous research by Park et al. (2005) and Ahari et al. (2015) highlighted the benefits of silica fume by a "ball bearing effect," which reduced yield stress and enhanced mobility when the SF content ranged from 5% to 10%. However, this effect diminished when the SF content exceeded 10%, primarily due to the heightened interparticle friction caused by the extremely fine particles of silica fume, as noted in the study by Arshad et al. (2021). A similar pattern was observed in the SF20 samples. At 20% SF content, the "ball-bearing" effect disappeared. Instead, adding silica fume increased the water demand of the mix, aligning with the findings of Correa-Yepes et al. (2018) and Memon et al. (2013).

Fig. 5 illustrates how the yield stress and plastic viscosity varied with different w/s ratios. The yield stress was significantly raised for 20% SF content, brought on by the higher water demand. However, a simultaneous increase in the targeted NaOH% lowered the yield stress value, suggesting a higher dissolution of the slag particles. In the case of the SF10 samples, the yield stress followed an order w/s = 0.50 > 0.55 > 0.45 (Fig. 5). However, the plastic viscosity followed w/s = 0.45 > 0.50 > 0.55. Thus, increasing the w/s ratio demonstrated a lower initial flow resistance, but once the material began to deform, it showed a higher internal friction, resulting in a higher viscosity. A higher interparticle friction indicated that the particles were bound more tightly, restricting the flow and resulting in a denser microstructure. The viscosity for w/s = 0.45 was high even if the yield stress was much lower, indicative of an ordered microstructure to have formed. The strength results, which will be covered in the upcoming sections, also confirmed this. For w/s = 0.55, a higher yield stress was followed by an immediate drop in viscosity. It is to be noted that all samples showed shear thinning behavior, i.e., $C/\mu_p < 0$.

The soda ash acted more as an accelerator when added as a premixed solution (Fig. 6), and a quick set or false set was observed for all premixed (PM) set-C mixes. The dissolution was subdued by the readily available $CO_3^{2-}$ ions in solution form. According to earlier studies (Garcia-Lodeiro et al. 2015), the presence of carbonate ions caused the medium to become more acidic and produce alkaline carbonates, which reduced the degree of dissolution (Fernández-Jiménez and Puertas 2001). On the other hand, in the set-A mixes, the soda ash first reacted with the hydrated lime to release $Na^+$, $OH^-$, $Ca^{2+}$, and $CO_3^{2-}$ ions; the $Ca^{2+}$ ions from hydrated lime exhausted the $CO_3^{2-}$ ions so that the $Na^+$ could potentially react with slag particles, lowering the yield stress. Also, anhydrous soda ash typically contained a higher alkali concentration than a soda ash solution of the same weight. Thus, the formation of the hydration product was masked by the reduced alkali concentration in the set-C mixes. In the control set-B mixes, on the other hand, the higher alkalinity of the NaOH solution raised the dissolution rate of the slag. The plastic viscosity and the yield stress were observed to have significantly decreased, making it clear that the gel structures formed unimpeded. This, in turn, made the mixture more fluid.

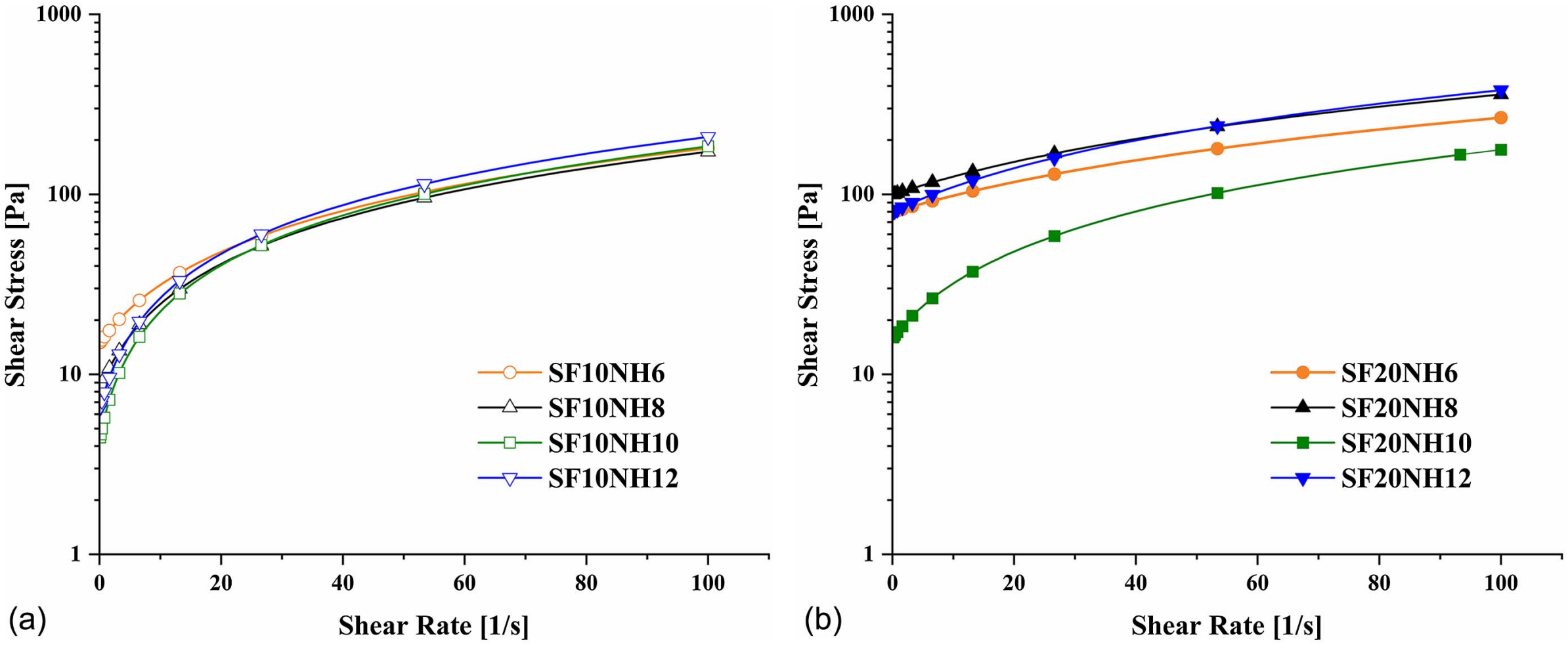

**Fig. 4.** Shear stress variation with the shear rate for different one-part pastes at w/s of 0.45 in (a) SF10 mixes; and (b) SF20 mixes.

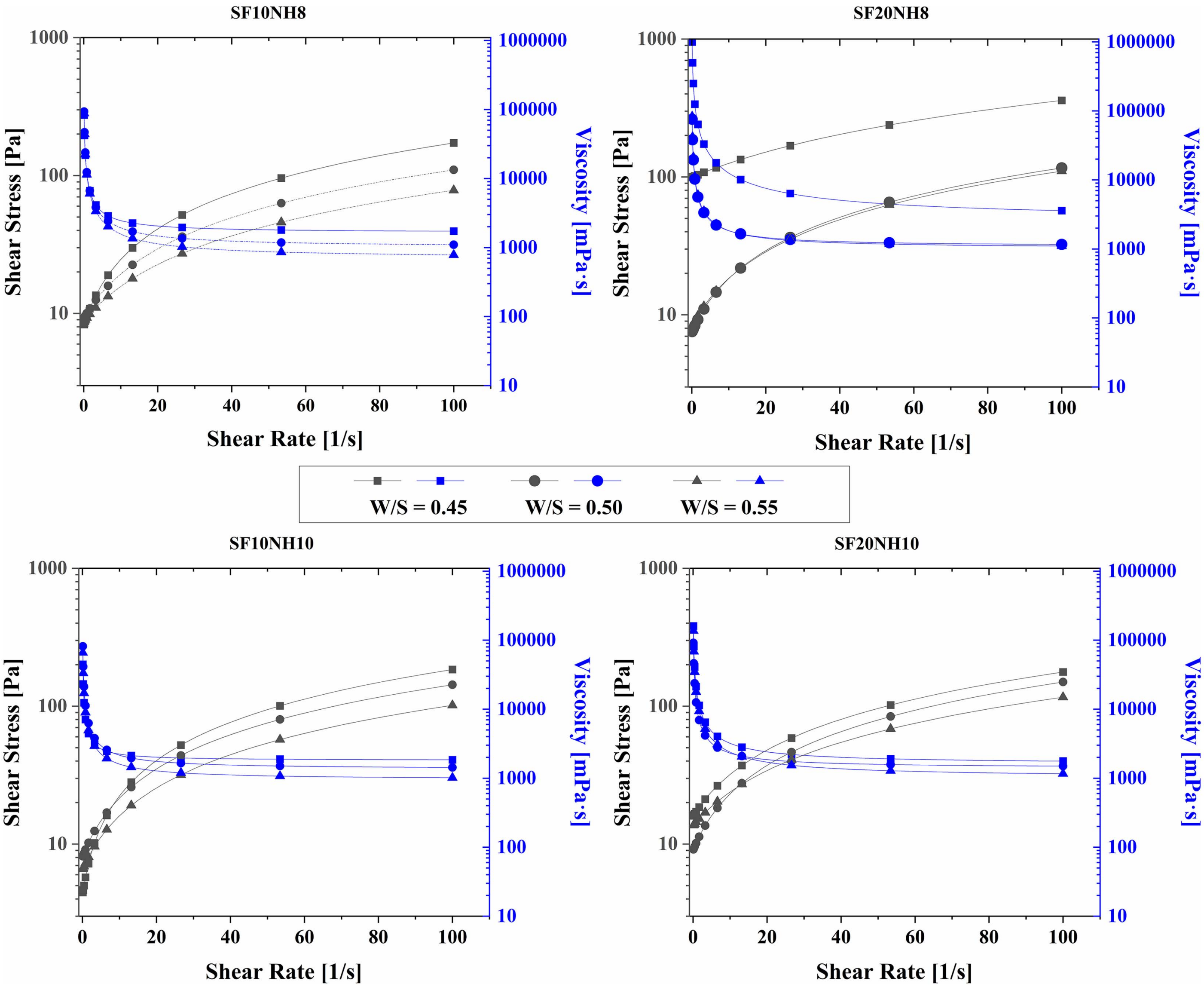

**Fig. 5.** Shear stress and viscosity variation of different one-part pastes at different w/s ratios.

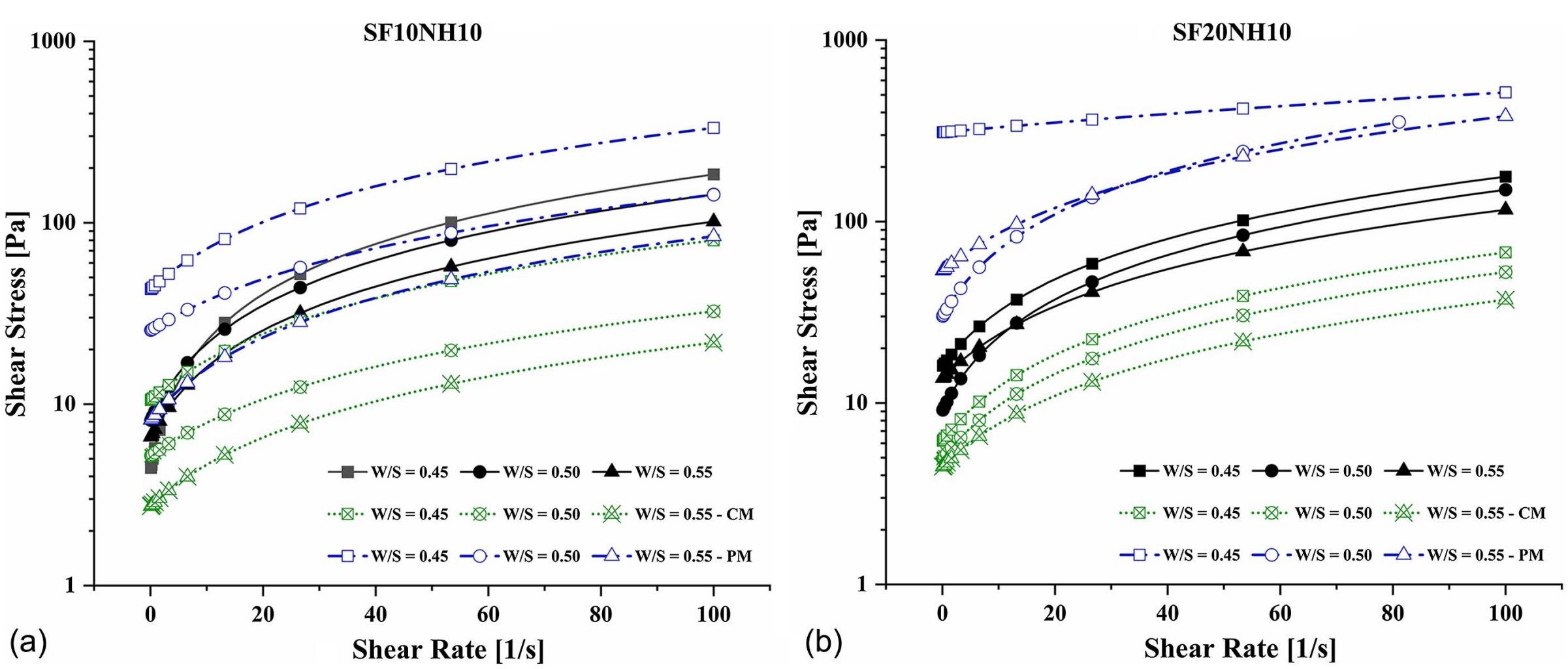

**Fig. 6.** Shear stress variation of (a) SF10NH10; and (b) SF20NH10 at different w/s ratios for different mixing procedures.

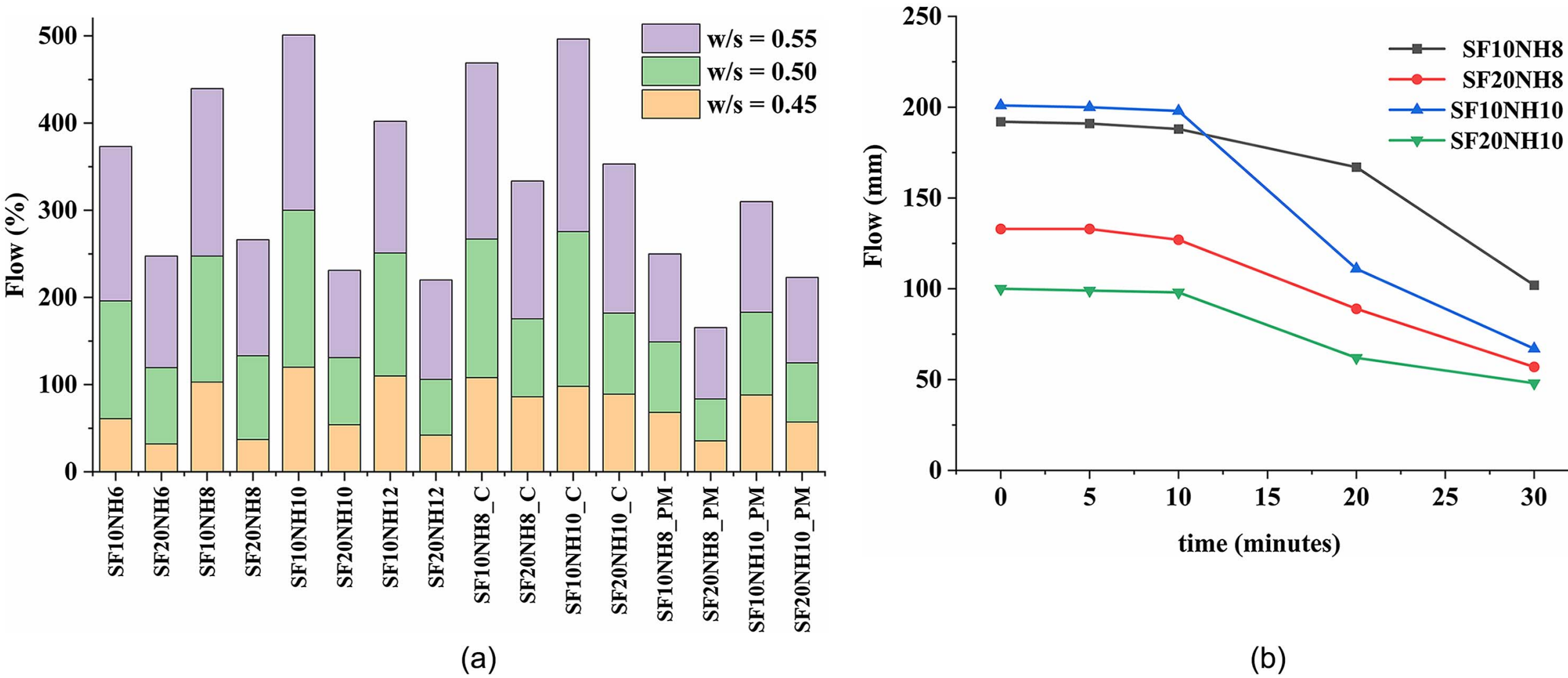

**Fig. 7.** (a) Flow values of the mixes at different w/s ratio; and (b) flow values of selected mixes at w/s = 0.55 at different time intervals.

### Mortar Flow

The flow spread diameters are plotted in Fig. 7(a). The premixed samples from set C exhibited lower flow compared to the set-A mixtures. In contrast, the set-B control mixtures had a greater spread, which can be attributed to the higher alkalinity of the NaOH solution, aligning with the observations made in terms of rheology. When the targeted NaOH content was increased, it led to a substantial spread for the set-A mixtures. Furthermore, increasing the SF content did not boost the flow of the mix, in line with the rheological observations and the findings from Correa-Yepes et al. (2018).

The chosen set-A mixtures were evaluated for their loss of fluidity at different time intervals, as depicted in Fig. 7(b). It was observed that the samples nearly lost their fluidity at approximately 30 min, indicating a rapid gel formation process. Among these samples, SF10NH8 exhibited the highest flowability at the 30-min mark compared to the others. This could be attributed to a slower hydration reaction, which can be linked to the lower targeted amounts of NaOH and SF. These findings demonstrate that flowability varied and was significantly influenced by the activator content, which in turn affected the solubility of aluminosilicates and the resulting reaction products.

### Compressive Strength

Fig. 8 portrays the strength development of the samples after 7, 28, and 120 days of curing. SF10NH8, SF20NH8, SF10NH10, and SF20NH10 have higher initial strengths. The 120-day strength of all samples reached a maximum in the same range for both 10% and 20% SF addition. Nevertheless, the rate of strength gain varied significantly depending on the percentage of SF added, suggesting the possibility of a secondary reaction. The initial strength was observed to be more than 20 MPa for a targeted NaOH percentage higher than 6%. In comparison, the later-age strength was more than 30 MPa, establishing the competency of the mixes for structural application. SF10NH8, SF20NH8, SF10NH10, and SF20NH10 demonstrated the best strength development, with a 28-day strength of 35.1 MPa and a 120-day strength of 41.33 MPa for SF10NH10. Hence, these four samples were selected as demonstrative samples to further explore the hydrated products' characteristics. Fig. 9 compares the strength development of the mixes for other w/s ratios. The compressive strength was reduced as the w/s ratio increased, as a higher w/s ratio tended to dilute the system by affecting the $H_2O/Na_2O$ ratio. As a result, ion dissolution was hindered, resulting in decreased strength.

The strength development was comparable except for SF20NH10_C for the two-part set-B mixes. Incidentally, the initial strength for all four blends was virtually the same. However, for SF20NH10_C, the later age strength failed to develop. This inconsistency could be due to the $Na^+$ and $Ca^{2+}$ ions. Since GGBFS was solely the source of $Ca^{2+}$ ions in the set-B mixes, a rise in NaOH at the same time as a decrease in GGBFS caused competition between the metal ions (Peyne et al. 2017) that delayed further reaction. However, the identical set of ions did not compete in the set-A mixes because a targeted incorporation of NaOH led to the simultaneous formation of $CaCO_3$, which engaged the $Ca^{2+}$ ions. This was consistent with the findings of Ma et al. (2019), who found calcite formation to be a stable polymerization product in a $Na_2CO_3$-activated fly ash mixture. Calcite filling was reported to be beneficial with reduced porosity (Wang et al. 2018).

The initial strength development was generally high for the premixed set-C samples (Fig. 10), in line with the previous studies (Akturk et al. 2019). In this case, the system of GGBFS-SF-HL was activated by soda-ash solution, and the hydrated lime served as an additive. According to earlier research (Fernández-Jiménez and Puertas 2001), the dissolution process was slow when $Na_2CO_3$ activated the binders. Calcite and sodium calcium carbonates formed when the abundant $CO_3^{2-}$ ions in the pore solution interacted with the Ca-O linkages in the slag and other calcium-rich compounds. Bernal et al. (2014) also reported slow activation by $Na_2CO_3$, where the primary reaction formed gaylussite. In the premixed mixes, the presence of $Na^+$ and $CO_3^{2-}$ readily exhausted the $Ca^{2+}$ from the HL to form gaylussite with little $Na^+$ left to activate the GGBFS. This delayed the subsequent reaction of GGBFS. This phenomenon likely contributed to the notable difference in strength between SF10NH8_PM and SF20NH8_PM. Also, in SF20NH8_PM, the higher SF content could readily react with $Ca^{2+}$ ions from GGBFS, leading to the formation of CSH, resulting in a significant increase in initial strength. Surprisingly, this difference was less pronounced in the NH10_PM mixes. SF10NH10_PM and SF20NH10_PM exhibited greater strength despite varying SF contents. The focus should be on the rate of strength gain in this context. While SF10NH8_PM and SF20NH8_PM showed limited strength improvement over 28 days, the NH10_PM samples displayed a considerable increase during the same period. This suggests that the availability of slag particles played a crucial role in 28-day strength. It became evident that more slag was accessible for further reaction in the NH10 PM samples, indicating ongoing polymerization and pozzolanic reactions. Therefore, higher $Na^+$ content in a premixed mixture appeared advantageous for strength development.

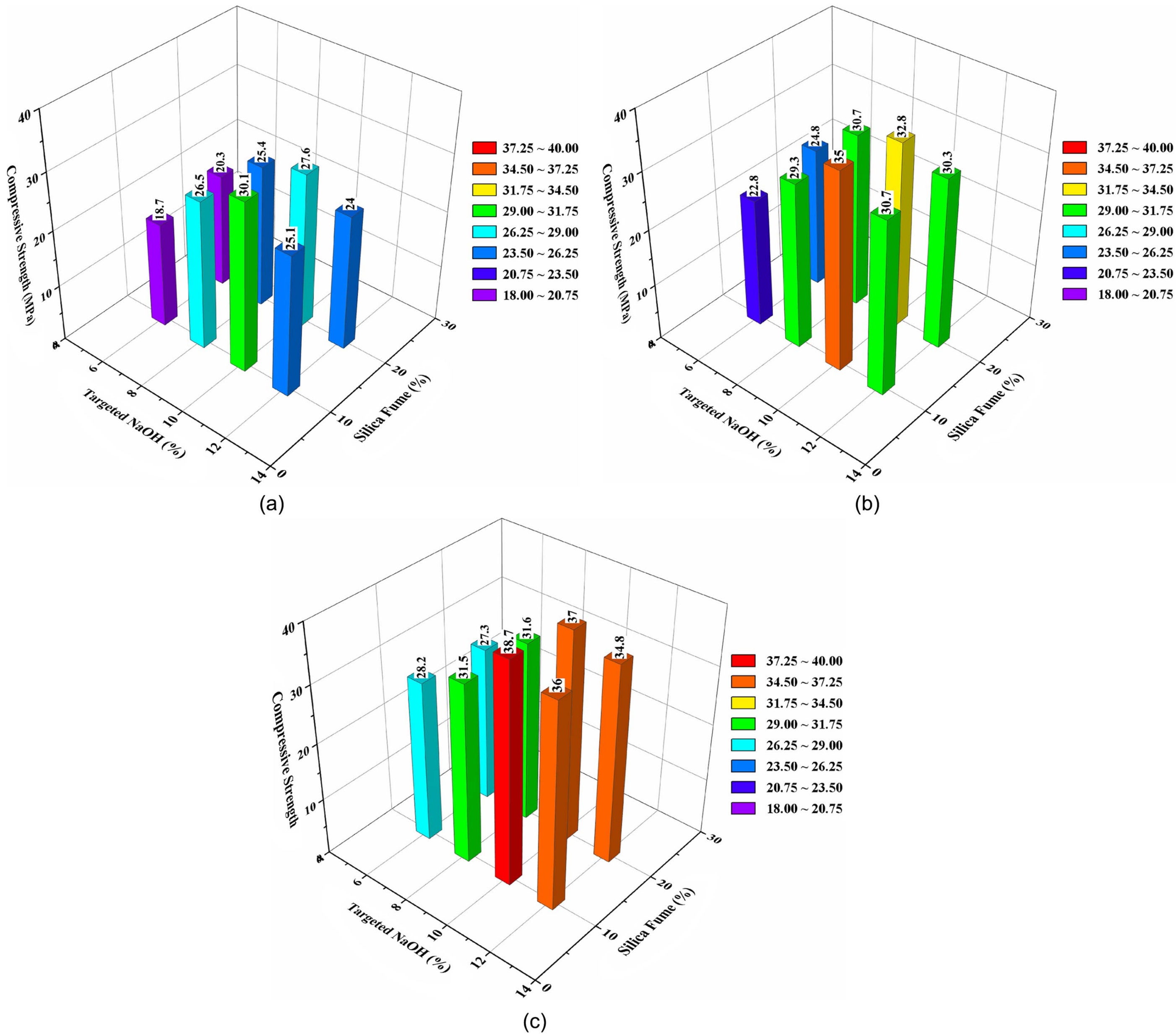

**Fig. 8.** Strength development of the Set-A mixes at w/s of 0.45 at different curing durations in the order of (a) 7 days; (b) 28 days; and (c) 120 days.

The same raw materials were the subject of an investigation by Kearsley et al. (2015), who reported compressive strengths of 45–50 MPa at 28 days and 6% alkali concentration. In the said study, the mechano-chemical processing of the raw materials did, however, help with the development of strength. The analytical chemicals' ability to attain high strength at a considerably lower content was also evident. The GGBFS used in the above study was reported to be neutral; it did not explicitly state the chemical grade, and the mix design was different, so it could not be used as a benchmark for comparing the strength development in the present study by the analytical grades. In order to validate any differences, additional tests were conducted to compare mixtures with analytical-grade $Na_2CO_3$ and $Ca(OH)_2$ complements.

A similar trend was observed for both industrial and analytical grades: SF10NH8 > SF20NH8 and SF10NH10 > SF20NH10 (Fig. 10). An excess of SF seemed to have hampered the process of alkali activation, as was established by Cheah et al. (2019). Additionally, the increased surface area of SF increased the overall water demand. Therefore, a further increase beyond the recommended content might result in inadequate wetting of the slag particles, delaying their breakdown (Khater 2013). When comparing them to each other, the analytical-grade chemical activators exhibited higher strength in the cases of SF10NH8 and SF10NH10. Conversely, they resulted in lower strength for SF20NH8 and SF20NH10, with SF20NH10 showing a particularly significant decrease. The finer reagents indeed enhanced reactivity, but the increased fineness of the solid activators, combined with SF, led to a higher water demand. This, in turn, imposed a water deficiency for the slag particles, impeding the dissolution, ultimately reducing strength. The higher the SF content, higher the deficiency. Consequently, a noticeable decline in strength was observed for both SF20NH8 and SF20NH10 in Fig. 10.

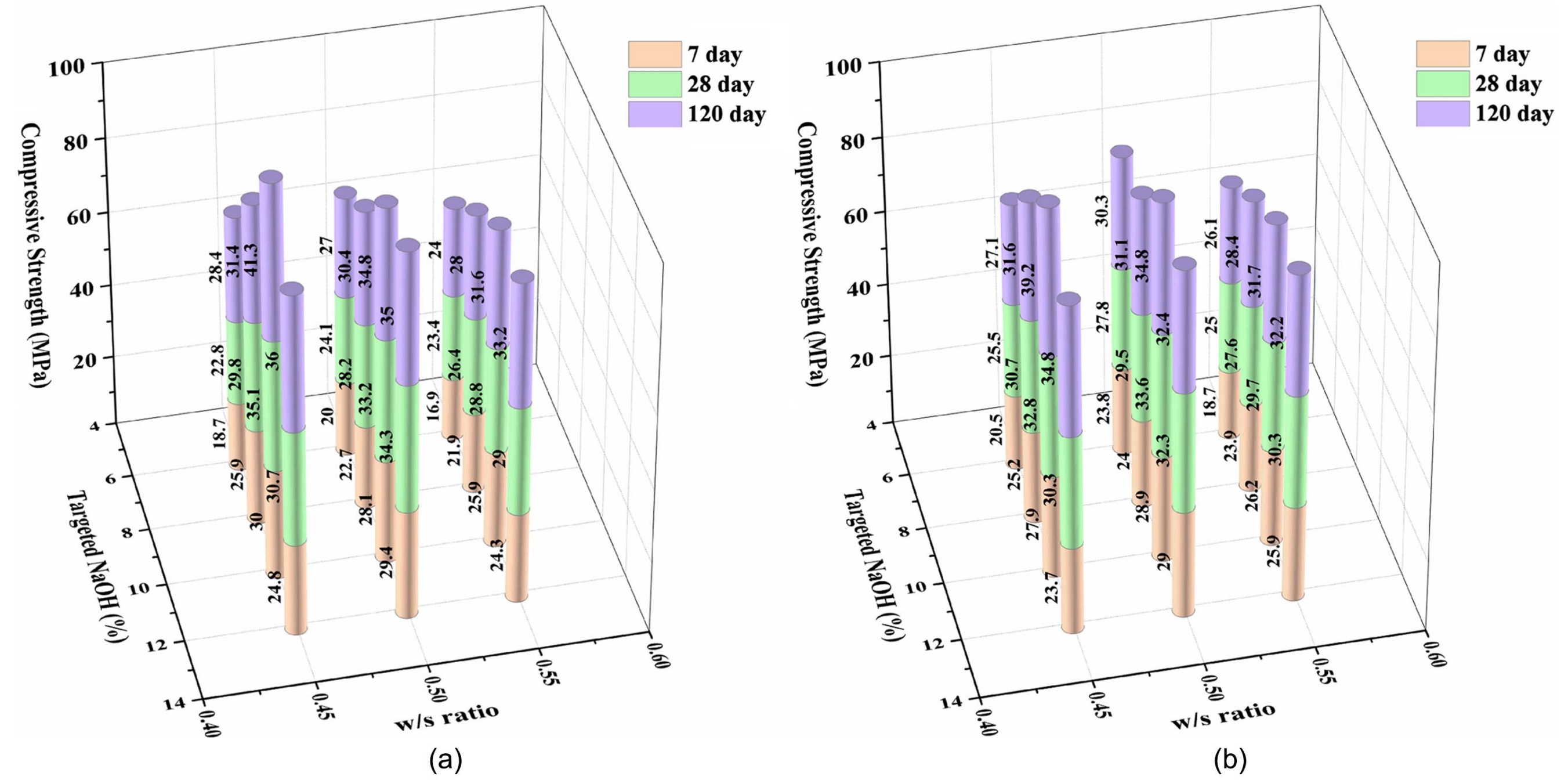

**Fig. 9.** Strength comparison of set-A mixes with different w/s ratios of (a) 10% SF addition; and (b) 20% SF addition.

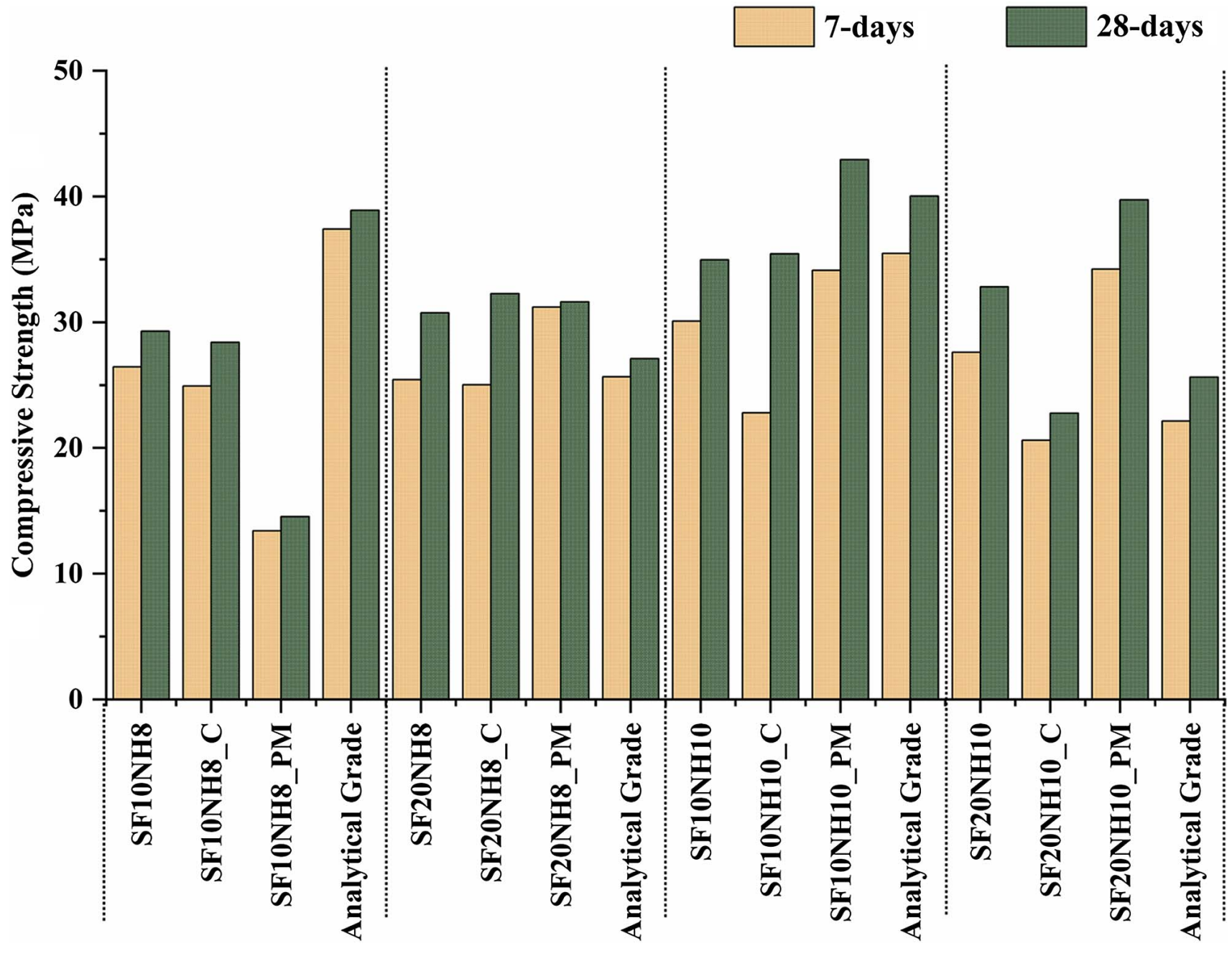

**Fig. 10.** Strength comparison of different set of mixes at w/s = 0.45 at different curing ages.

## Phase Detection through X-Ray Diffraction

Since they had the maximum strength at w/s 0.45, SF10NH8, SF20NH8, SF10NH10, and SF20NH10 were chosen as demonstrative samples to characterize and study the hydrated products. As subjects of the current investigation, Fig. 11 displays the XRD patterns of the selected set-A samples at 7 and 28 days. At the age of 7 days, XRD analysis identified CSH, CASH, CNASH (#98-002-2602), and NASH (#98-001-2576) as the primary binding phases. The mineral phases of thaumasite and hydrotalcite were also found in the 7-day samples, although calcite (#00-005-0586), magnesites and dolomite were the dominant carbonate phases in the 28-day samples. All samples at 7 days identified akermenite-gehlenite at 31.14° ($2\theta$), representative of unreacted GGBFS, hence suggesting the gradual dissolution of aluminosilicates.

In the 7-day samples, calcite peaks were apparent at 22.9°, 35.9°, 43.3°, 47.0°, 57.5°, and 60.9° ($2\theta$). Other major calcite peaks were seen at 29.4°, 39.5°, 43.3°, and 48.4° ($2\theta$), which minored at 28 days. The calcite formation resulted from the primary cation exchange in Eq. (1). The pronounced calcite peak at 39.5° ($2\theta$) overlapped with a CSH peak on a background convexity owing to the semiamorphous nature of the latter, as reported by Akturk et al. (2019) and Walkley et al. (2017). NASH peaks were visible at 26.7°, 42.5°, and 48.9° ($2\theta$), which morphed into the convexity at 28 days. The intensity of calcite peaks decreased at 28 days due to the consequent formation of magnesites and dolomites. Diffraction peaks of magnesite were specifically at 32.9°, 43.2°, and 36.4° ($2\theta$) for 28-day samples. Characteristic peaks of dolomites were at 60.0° and 67.6° ($2\theta$), overlapping with quartz, which increased in intensity at 28 days. Minor hydrotalcite peaks were visible in the samples at 39.3° and 75.7° ($2\theta$), which increased in intensity at 28 days, aligning with previous works of literature (Akturk et al. 2019; Walkley et al. 2017). SF10NH10 and SF20NH10 prominently showed the major peak of hydrotalcite at 11.3° ($2\theta$) at 7 days.

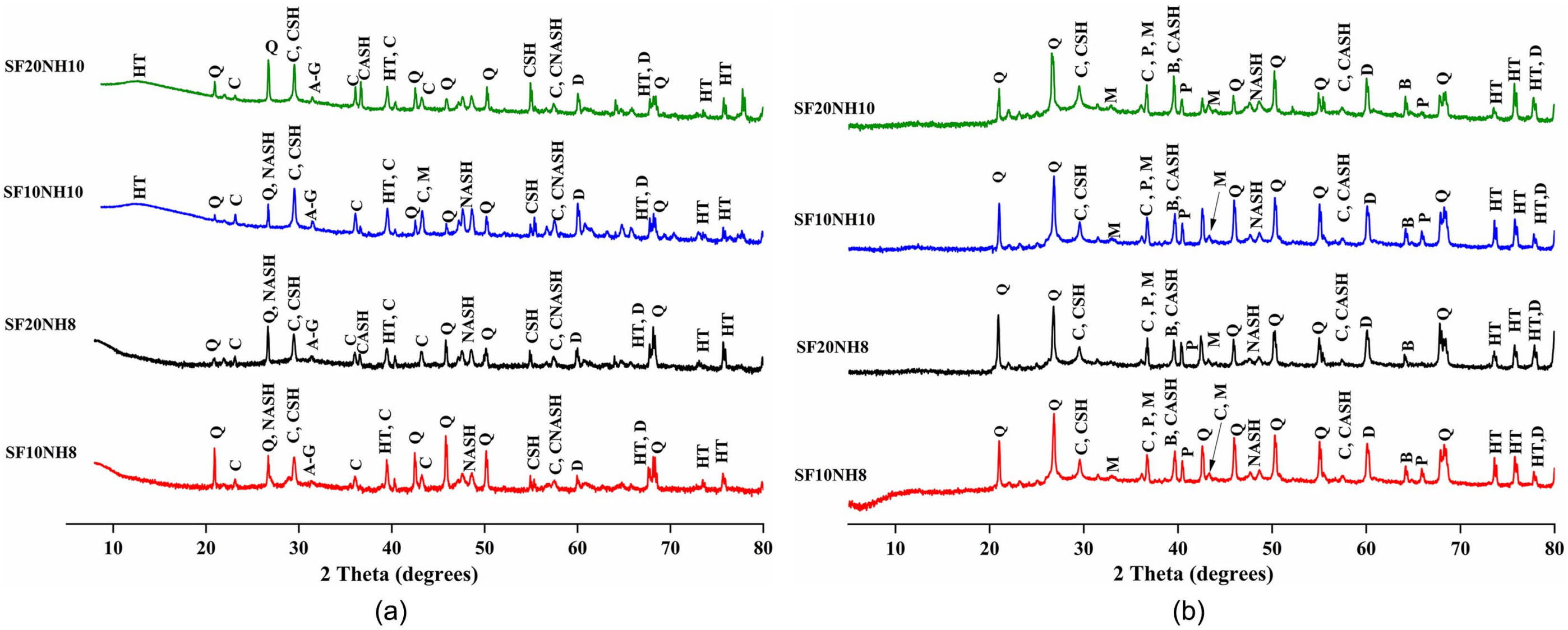

**Fig. 11.** XRD of the mortar samples at (a) 7 days; and (b) 28 days. Q = quartz; P = Portlandite; C = calcite; B = brucite; M = magnesite; D = dolomite; and HT = hydrotalcite.

Sharp diffraction peaks are generally replaced by a broadly dispersed halo when x rays are diffracted on an amorphous section. So, a convexity cantered at around 21°–37° ($2\theta$) with minor peaks in all four samples at 28 days. This broad hump was the usual peak of amorphous gels, such as CSH (Haha et al. 2011), CASH (Walkley et al. 2017), NASH, and CNASH (Bakharev 2005; De Silva et al. 2007) gels, indicative of the diffraction pattern of alkali-activated materials; the more convex, the higher the strength. The XRD analysis helped to clarify how the set-A mixtures' strengths developed. Although the carbonate phases, viz., magnesites and dolomites, congealed, the increase in the convexity of the hump asserted the gel phases' formation. Eventually, this led to a rise in the strength. It is important to note that whilst the 7-day strength was solely attributable to CSH, CASH, or NASH, and CNASH, the 28-day strength resulted from the coexistence of CSH, CASH, and NASH (Fig. 11). The presence of hydrotalcite also influenced the initial strength. Reactive MgO addition to GGBFS blends was reported to have increased mechanical strength through hydrotalcite formation (Adeleke et al. 2021; Haha et al. 2011). The chemistry of the CNASH gel was greatly influenced by the amount of $Mg^{2+}$ available (Walkley et al. 2017), which accounted for CNASH's dominance during the early stages of strength growth. Increasing the concentration of $Mg^{2+}$ ions resulted in higher hydrotalcite phases in the material (Gu et al. 2014; Jin et al. 2015). However, the presence of akermanite–gehlenite phases can impede the formation of hydrotalcite phases, allowing more Al to be available for substitution within a CNASH framework. With an adequate amount of Ca, though, an increase in Mg content promoted the formation of low-Al CASH phases and portlandite, as the hydrotalcite phases formed, consequently reducing the Al content in the material. This clarified the high hydrotalcite and minor portlandite peaks in Fig. 11(b). Brucite peaks were also apparent at 39.6° and 64.5° ($2\theta$). The CNASH gels were eventually destabilized to form CASH gels (Garcia-Lodeiro et al. 2011). As the dissolution of GGBFS progressed, the NASH gels formed and coexisted with the CASH gels.

In complex crystal structures, especially when diffraction peaks overlap, it might not be easy to distinguish precisely between signals from different crystallographic locations or phases. Quantifying the existence of multiple phases or the presence of amorphous material in a sample can be a demanding task and might necessitate the application of additional analytical methods. Though Rietveld refinement is primarily tailored for examining crystalline materials and might not be well-suited for analyzing substances that lack an ordered atomic structure, it might provide an overall approximation of the phases present. Hence, Table 5 tabulates an approximation of the phases, obtained by Rietveld refinement. SF10NH10 showed the highest content of NASH (7.19%), followed by SF10NH8 (4.46%), while SF20NH10 and SF20NH8 showed the highest CASH content, 2.4% and 2.08%, respectively. Thus, higher SF content led to higher formation of CASH, while lower SF content led to higher content of NASH. Table 5 shows that CSH and NASH were responsible for the strength development in SF10NH10, whereas CSH, CASH, and NASH were responsible for the growth in SF20NH10. The NH8 samples showed a similar pattern. SEM and TG-DTA were used to confirm further the observations made in XRD.

### Microstructural Characterization

Scanning electron microscopy (SEM), coupled with the energy dispersive spectroscopy (EDS) technique, as shown in Figs. 12 and 13, was done to map and ascertain the elemental composition of the selected set-A mixes. All four samples initially showed substantial quantities of cubic crystal precipitate of calcite. The primary ion exchange led to calcite formation [Eq. (1)]. However, the deposition of the CSH, CASH, and NASH gels obscured the presence of calcite in the 28-day samples. Initially, the microstructure exhibited unreacted GGBFS, much like XRD had shown akermenite–gehlenite. However, it transformed into compact, dense systems after 28 days, consistent with the strength findings. SF10NH10 exhibited the presence of hydrotalcite deposits, whereas SF20NH10 displayed the formation of aggregated calcite.

**Table 5.** Approximate quantities of the phases formed at 7 and 28 days through Rietveld refinement

| Phases | SF10NH8 | | SF20NH8 | | SF10NH8 | | SF20NH10 | |
|---|---|---|---|---|---|---|---|---|
| | 7-day | 28-day | 7-day | 28-day | 7-day | 28-day | 7-day | 28-day |
| Hydrotalcite | 1.00 | 3.41 | 0.06 | 1.33 | 11.59 | 2.91 | 6.21 | 1.37 |
| Brucite | 0 | 2.1 | 0 | 0.9 | 0 | 1.85 | 0 | 0.09 |
| Calcite | 68.23 | 16.53 | 70.53 | 24.39 | 59.53 | 13.83 | 89.74 | 29.7 |
| Portlandite | 0 | 2.21 | 0 | 0.67 | 0.03 | 3.13 | 0 | 0.88 |
| Magnesite | 0 | 9.49 | 0 | 6.37 | 0 | 9.00 | 0 | 7.20 |
| Dolomite | 0.46 | 4.92 | 1.55 | 4.23 | 3.27 | 3.79 | 0 | 4.95 |
| Akermenite-Gehlenite | 1.35 | — | 0.68 | — | 0.63 | — | 0 | — |
| Tobermorite | 21.32 | 56.18 | 17.5 | 55.56 | 20.08 | 55.26 | 0 | 50.82 |
| Zeolite X (CASH) | 0 | 0.41 | 0.45 | 2.08 | 0 | 0.91 | 2.4 | 2.4 |
| Zeolite X (NASH) | 2.77 | 4.46 | 2.63 | 3.38 | 3.41 | 7.19 | 0 | 2.28 |
| Zeolite A (CNASH) | 4.95 | 0.31 | 6.64 | 1.18 | 1.5 | 2.12 | 1.65 | 0.82 |

The EDS spectra showed the composition was abundant in Ca and Si, with moderate contributions from Na, Mg, and Al. The chief elements influenced CSH gel formation, while Ca, Na, Si, and Al were responsible for CASH and NASH formation; Mg contributed to hydrotalcite, magnesite, and dolomite formation. Table 6 shows the EDS data for an average of 20 points. SF20NH10 showed a lower Na, Al, and Mg concentration, suggesting that the alkali did not aid in the initial dissolution of slag. In other words, the hydration reaction was occurring more slowly in SF20NH10. The XRD examination revealed that SF20NH10 had more calcite initially. This made the SF react readily with the Ca and Al imparted by GGBFS. Hence, the initial CASH synthesis might be the only reason for the 7-day strength of SF20NH10. For SF10NH10, however, it was clear that CSH, NASH, and CNASH all had an impact on the initial strength, as determined by the XRD analysis. Additionally, by 28 days, the Al concentration in SF10NH10 shows a more noticeable drop, indicating generous precursor (being the only source) engagement.

An increased calcium content $[Ca/(Si + Al) \rightarrow 1]$ resulted in the formation of CNASH and secondary products, such as portlandite, along with a CASH gel that had a moderate level of Si (Walkley et al. 2017). Conversely, a lower calcium content $[Ca/(Si + Al) \approx 0.65]$ led to higher levels of NASH and CASH with reduced silicon content. This pattern is also observed in Table 6. According to Jagad et al. (2023b) and Reddy et al. (2018), as the reactions progressed, $Ca/(Si + Al)$, Ca/Na, and Ca/Si declined, while $Na/(Si + Al)$ increased. The trends observed in the molar ratios for both 7 and 28 days were consistent. It is worth noting that the $Na/(Si + Al)$ ratio did not show a significant increase for SF20NH8 over the course of 28 days. However, the $Ca/(Si + Al)$ ratio decreased significantly, accompanied by a reduction in the Ca/Na and Ca/Si ratio. This suggested that the increase in strength observed at 28 days was likely due to a rise in CSH and CASH, with minimal to no increase in NASH and CNASH. In CNASH gel, the alkali $Na^{+}$ on the outside endffered anion binding mechanisms (Walkley et al. 2017). So, Na served as a charge modifier in the negatively charged Al-O-Si system. The CNASH destabilized to CASH, with the declining Ca/Na ratio supporting the decreasing CNASH as seen in the XRD. It is worth mentioning that the lower Ca/Na values resulted from higher concentrations of $Na^{+}$ in NH10 mixes compared to NH8 mixes, as demonstrated in Table 6.

$Mg^{2+}$ in a $Ca(OH)_2$ or $CaCO_3$ system could form magnesium-containing carbonate and liberate additional $Ca^{2+}$ ions (De Silva et al. 2009). The XRD showed brucite and magnesite at 28 days, proportional to the Mg/Ca ratio. The Mg/Al ratio ranged between 0.2 and 0.5 for all samples, confirming the presence of hydrotalcite (Kovtun et al. 2015; Wang and Scrivener 1995). At room temperature, magnesium in an alkaline medium could precipitate (hydrated) magnesium carbonate (Montes-Hernandez et al. 2016). A release of water molecules stoichiometrically balances off the decrease in calcite peaks and subsequent magnesite and dolomite formation (equation). This accounted for SF10NH10 to have a higher amount of free water despite having higher strength. At 28 days, the Mg/Ca for SF10NH10 increased significantly, implying a higher content of magnesites. The molar ratios revealed by SEM-EDS agreed well with the XRD analysis. At 28 days, SF10NH10 gave a higher $Na/(Si + Al)$ ratio, followed by SF10NH8, as for Ca/Si, SF20NH10 and SF20NH8 nearly had the same content.

## Thermal Analysis

### Phase Transitions

Fig. 14 depicts the TG-DTG and DTA thermograms for the selected mixes at 28 days. Similar trends in mass loss were seen in all four samples. As seen in Fig. 14(b), substantial peaks and their descent back to the baseline represented various stages of decomposition. Visualizing the peaks and their drop in Fig. 14(b), seven occurrences of weight losses in the temperature ranges were grouped and noted as (1) 32°C to 105°C for free water removal ($Ldh_a$); (2) 105°C to 150°C for crystalline water removal ($Ldh_b$); (3) 150°C to 230°C for intralayer water removal ($Ldh_c$); (4) 230°C to 460°C for brucite and hydrotalcite dehydroxylation ($Ldx_a$); (5) 460°C to 530°C for portlandite dehydroxylation, ($Ldx_b$); (6) 530°C to 635°C for magnesite decomposition, ($LdC_m$); and (7) 635°C to 1,000°C for calcite and dolomite decomposition ($LdC_c$). The DTA thermogram further validated the phase transitions in the sample.

Free water evaporation was indicated by the mass loss below 105°C, the first endotherm in the DTA. The mass loss between 105°C and 150°C in the TG-DTG thermogram could be attributed to the removal of crystalized water from the hydrotalcite phase (Ponce-Antón et al. 2018), which was in good accord with XRD analyses. A higher mass loss would thus correspond to more hydrotalcite presence. The intralayer water removal from the hydrated phases, i.e., structural $OH^{-}$ ions breakdown, occurred between 150°C and 230°C. The corresponding DTA thermogram [Fig. 14(c)] showed a minor change in the slope at C, indicating the onset of water removal, notably for C-A-S-H type gels, at around 180°C. Beyond C, a strong exotherm was recorded, most likely due to successive water removal from the CSH at approximately 200°C.

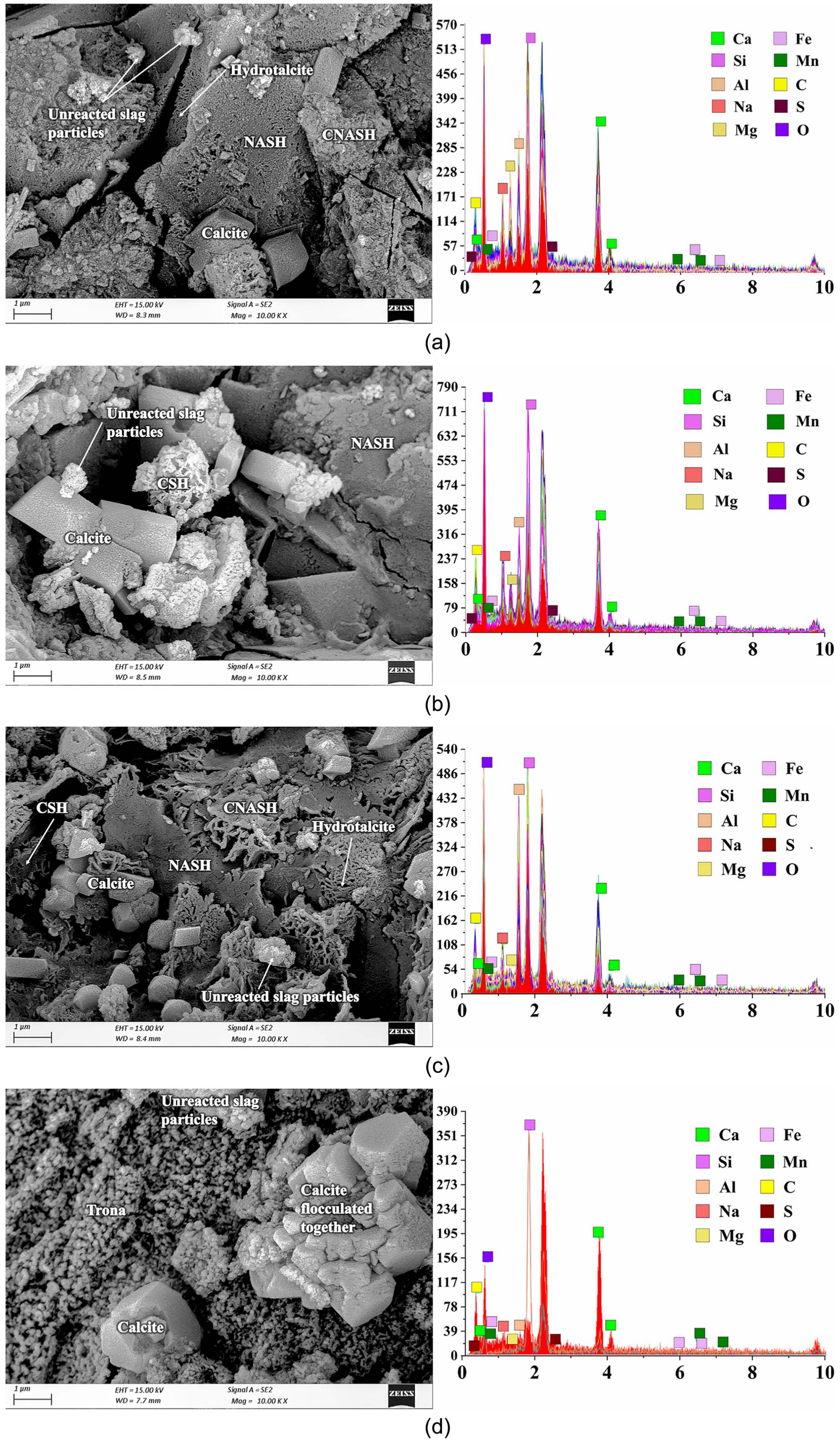

**Fig. 12.** SEM images and EDS spectra of mortar at 7 days of (a) SF10NH8; (b) SF20NH8; (c) SF10NH10; and (d) SF20NH10.

The dehydrated hydrotalcite's hydroxyl decomposition or the dehydroxylation occurs between 230°C and 460°C, followed by the subsequent formation of magnesium aluminate, magnesia, $CO_2$, and water. Brucite decomposition occurred in the same range. Hydrotalcite is a layered double hydroxide (LDH) with a high water content, with divalent and trivalent metal cations with a charge-compensating anion sandwiched between them. The loss of the $OH^-$ groups bound to $Al^{3+}$ and $Mg^{2+}$ of the Mg—Al—$CO_3$ compounds was responsible for the weight losses correlated with the decomposition of the LDH phases (Ponce-Antón et al. 2018; Yang et al. 2002). Over this range, brucite and portlandite also caused mass loss. XRD did detect minor peaks of brucite and portlandite in the 28-day samples. Portlandite dehydroxylation happened close to 500°C via CaO + $H_2O$, in line with (Song et al. 2018). Fig. 14(c) showed a minor hump between D and E, depicting the mass loss due to small amounts of portlandite. Hence, the mass loss, Ldx in Fig. 14(b), was owed to the breakdown of multiple phases—the $Mg_6Al_2(OH)_{16}CO_3$ phase, brucite and portlandite phase.

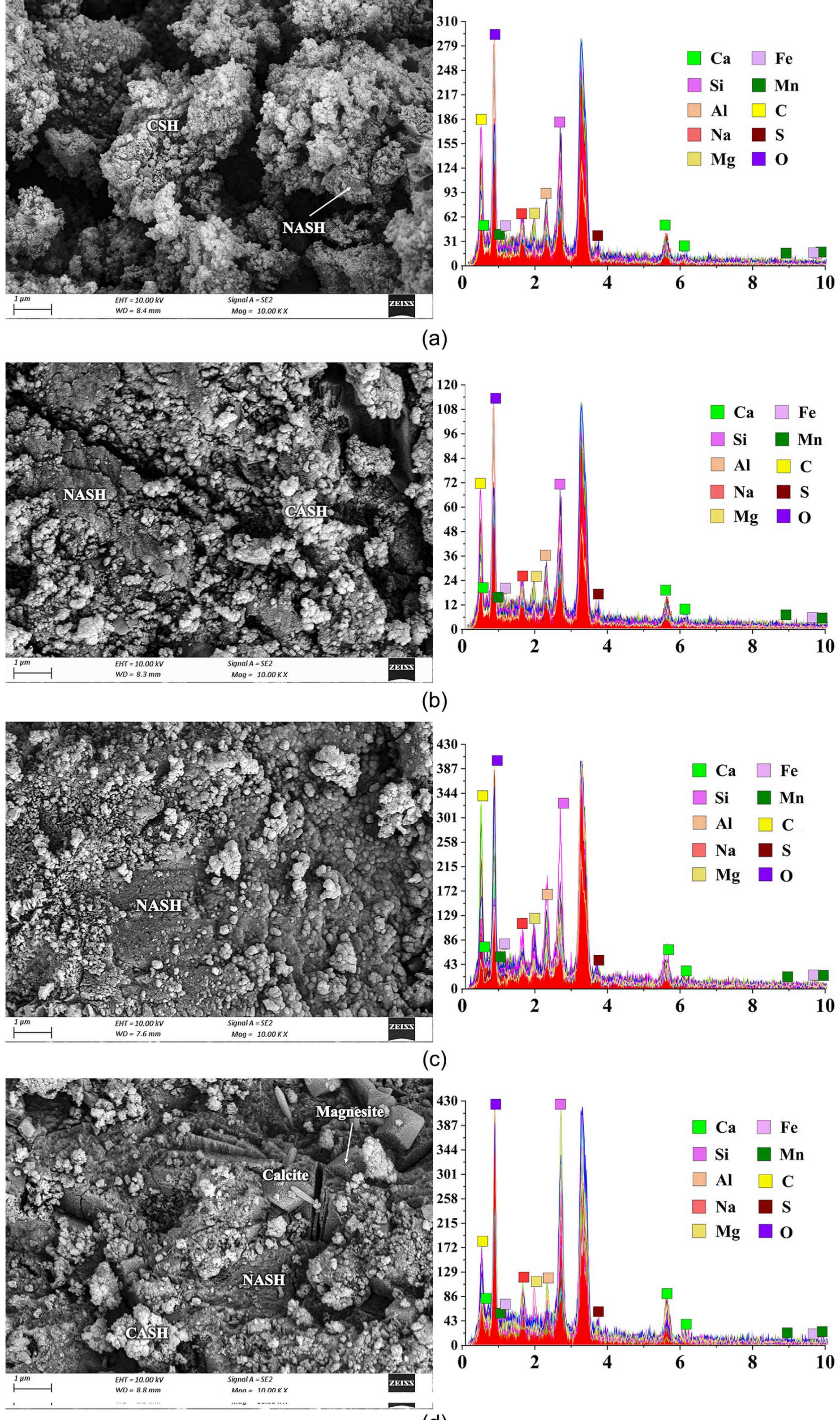

**Fig. 13.** SEM images and EDS spectra of mortar at 28 days of (a) SF10NH8; (b) SF20NH8; (c) SF10NH10; and (d) SF20NH10.

**Table 6.** Average elemental composition ratios acquired on selected samples by EDS

| Mix | Na | Mg | Al | Si | Ca | Ca/Si | Ca/(Si + Al) | Na/(Si + Al) | Mg/Ca | Mg/Al | Ca/Na | Si/Al | Compressive strength |
|---|---|---|---|---|---|---|---|---|---|---|---|---|---|
| 7 days | | | | | | | | | | | | | |
| SF10NH8 | 4.48 | 3.14 | 8.4 | 24.31 | 26.66 | 1.10 | 0.82 | 0.14 | 0.12 | 0.37 | 5.95 | 2.89 | 25.93 |
| SF20NH8 | 2.18 | 2.55 | 6.17 | 18.85 | 29.17 | 1.55 | 1.17 | 0.09 | 0.09 | 0.41 | 13.38 | 3.06 | 25.23 |
| SF10NH10 | 4.36 | 0.29 | 12.39 | 20.22 | 29 | 1.43 | 0.89 | 0.13 | 0.01 | 0.02 | 6.65 | 1.63 | 29.97 |
| SF20NH10 | 0.23 | 0.12 | 0.51 | 48.94 | 26.62 | 0.54 | 0.54 | 0.00 | 0.00 | 0.24 | 115.74 | 95.96 | 27.9 |
| 28 days | | | | | | | | | | | | | |
| SF10NH8 | 4.64 | 2.43 | 7.01 | 18.41 | 16.76 | 0.91 | 0.66 | 0.18 | 0.14 | 0.35 | 3.61 | 2.63 | 29.76 |
| SF20NH8 | 2.16 | 2.91 | 5.99 | 23.85 | 21.58 | 0.90 | 0.72 | 0.07 | 0.13 | 0.49 | 9.99 | 3.98 | 30.74 |
| SF10NH10 | 5.53 | 3.63 | 7.93 | 20.87 | 17.03 | 0.82 | 0.59 | 0.19 | 0.21 | 0.46 | 3.08 | 2.63 | 35.1 |
| SF20NH10 | 3.51 | 1.47 | 4.4 | 24.57 | 21.92 | 0.89 | 0.76 | 0.12 | 0.07 | 0.33 | 6.25 | 5.58 | 32.82 |

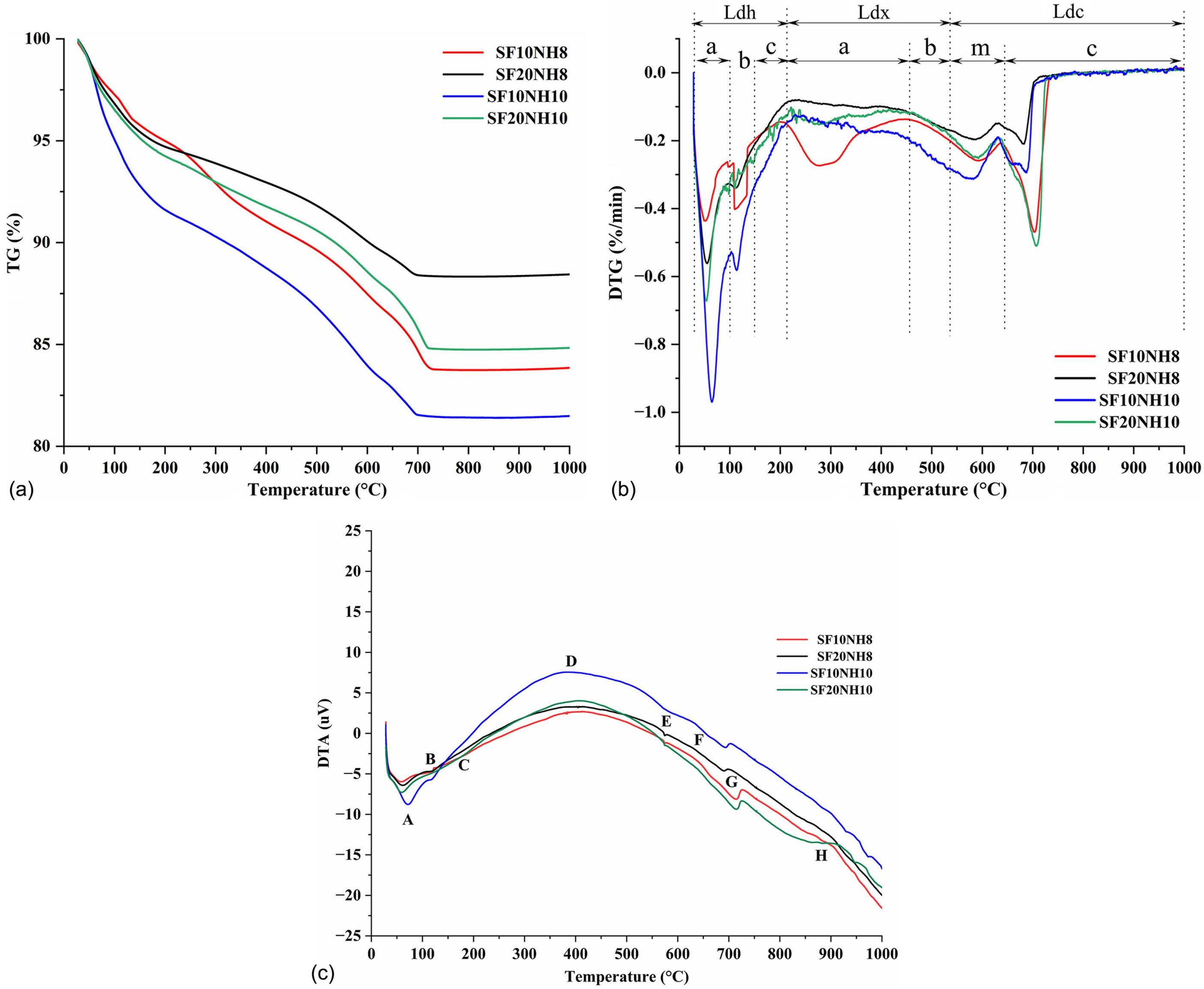

**Fig. 14.** (a) TG; (b) DTG; and (c) DTA curves of mortar at 28 days of SF10NH8, SF20NH8, SF10NH10, and SF20NH10.

According to reports, magnesite usually starts transforming at an average offset temperature of 467°C (Khan et al. 2001; Mahon et al. 2021). So, the $MgCO_3$ breakdown caused the DTG [Fig. 14(b)] to peak at around 588°C. The characteristic amorphous profiles of C─S─H began to change from 615°C to 630°C, and an upward exotherm was observed [at F, Fig. 14(c)] at around 635°C due to crystallization. A simultaneous endotherm–exotherm was observed between 690°C and 715°C, with the DTG curve peaking at 708°C likely due to dolomite decomposing into magnesia, lime, and carbon dioxide, with simultaneous calcite degeneration. An endothermic DTA peak was noted at E, characteristic of $\beta$-$\alpha$ quartz transformation of the sand.

### Significance of TG-DTA Results

Deboucha et al. (2017) introduced an empirical formula for indirectly measuring bound water in cementitious mixes using TG-DTA. Their research was built upon the previous studies by Bhatty (1986), Monteagudo et al. (2014), and Pane and Hansen (2005). They incorporated additional binders and mineral additives into their calculations of bound water. Since the raw materials used in this study were solid, Deboucha's formula could be effectively applied to quantify bound water and detect gel formation. Bhatty's approach (Bhatty 1986) considered mass loss between 105°C and 1,000°C. However, it did not account for carbonation caused by anhydrous materials ($Ldc_a$) when unreacted particles were present. Pane and Hansen (2005) adjusted for additional carbonation from anhydrous unreacted materials but did not include a conversion factor for portlandite carbonation. Monteagudo's method (Monteagudo et al. 2014) combined both aspects, yielding a unified expression for portlandite and anhydrous material. In Deboucha et al. (2017), the influence of mineral additives was incorporated, considering two significant effects: the filler and chemical effects and mass drift. In the present study, Deboucha's method underwent slight modifications to estimate bound water associated with each selected sample while accounting for the impact of additives on hydration kinetics. Chemical activators in solid form played a role in nucleating precursor grains, although device drift was not considered due to the minimal heating of the crucible:

$$W_B = (Ldh_b + Ldh_c) + (Ldx_a + Ldx_b) + 0.41LdC_m + 0.41(LdC_c - Ldc_a) - \mathrm{K} \tag{2a}$$

$$Ldh_a = \frac{W_{32^\circ} - W_{105^\circ}}{W_{32^\circ}} \times 100; \quad Ldh_b = \frac{W_{105^\circ} - W_{150^\circ}}{W_{105^\circ}} \times 100; \quad Ldh_c = \frac{W_{150^\circ} - W_{230^\circ}}{W_{105^\circ}} \times 100 \tag{2b}$$

$$Ldx_a = \frac{W_{230^\circ} - W_{460^\circ}}{W_{105^\circ}} \times 100; \quad Ldx_b = \frac{W_{460^\circ} - W_{530^\circ}}{W_{105^\circ}} \times 100 \tag{2c}$$

$$LdC_m = \frac{W_{530^\circ} - W_{635^\circ}}{W_{105^\circ}} \times 100; \quad LdC_c = \frac{W_{635^\circ} - W_{1000^\circ}}{W_{105^\circ}} \times 100 \tag{2d}$$

$$K = (m_{\text{precursor}} \times LOI_{\text{precursor}} + m_{\text{additive}} \times (LOI_{SF} + LOI_{SA} + LOI_{HL})) \tag{3}$$

$$m_{\text{precursor}} = \frac{m_s - m_b(x_{SF} + x_{SA} + x_{HL} + \frac{W}{B})}{1 + LOI_{\text{precursor}}}; \quad m_{\text{additive}} = \frac{m_s - m_b(x_{\text{GGBFS}} + \frac{W}{B})}{1 + LOI_{SF} + LOI_{SA} + LOI_{HL}} \tag{4}$$

$m_s$, $m_b$, $x_{SF}$, $x_{SA}$, $x_{HL}$, $W/B$, $LOI_{SF}$, $LOI_{SA}$, and $LOI_{HL}$ are the initial mass of the sample and total binder mass, all considered as the equivalent of 1 g of the sample, the replacement level of the additives (mineral and chemical), the water/binder ratio, and the loss on ignition of the additives (mineral and chemical), respectively. The bound water was evaluated by deducting the mass loss due to the decarbonation of all carbonate phases accounted by ($\mathrm{Ldc}$–$\mathrm{Ldc_a}$), as indicated in Table 7. The total correction factors are explained in Eqs. (5)–(8):

$$\mathrm{Ca(OH)_2 + CO_2 \rightarrow CaCO_3 + H_2O,} \quad \mathrm{CaCO_3 \rightarrow CaO + CO_2 \uparrow} \tag{5}$$

$$W_{B\,\text{carbonated}\,CH} = \frac{W_{m_{\mathrm{H_2O}}}}{W_{m_{\mathrm{CO_2}}}} \times \mathrm{Ldc} = 0.41\,\mathrm{LdC_c} \tag{6}$$

$$\mathrm{Mg(OH)_2 + CO_2 \rightarrow MgCO_3 + H_2O,} \quad \mathrm{MgCO_3 \rightarrow MgO + CO_2 \uparrow} \tag{7}$$

$$W_{B\,\text{carbonated}\,MH} = \frac{W_{m_{\mathrm{H_2O}}}}{W_{m_{\mathrm{CO_2}}}} \times \mathrm{Ldc} = 0.41\,\mathrm{LdC_m} \tag{8}$$

$$\mathrm{Mg_6Al_2(OH)_{16}CO_3 \rightarrow MgAl_2O_4 + 5MgO + CO_2 \uparrow + 8H_2O \uparrow} \tag{9}$$

As shown in Eq. (10), an adjustment factor was necessary to consider carbonated portlandite and calcite formation through ion exchange when calculating the portlandite content (CH):

$$\mathrm{CH} = \left(\frac{74.09}{18.01}\right)\mathrm{Ldx_b} + \left(\frac{74.09}{44.01} - \%_{\text{calcite}}\right)(\mathrm{LdC_c} - \mathrm{Ldc_a}) \tag{10}$$

where $(74.09/18.01)$ = molar mass ratio of $\mathrm{Ca(OH)_2}$ to water; $(74.09/44.01)$ = molar mass ratio of $\mathrm{Ca(OH)_2}$ to $\mathrm{CO_2}$; and $\%_{\text{calcite}}$ = weight fraction of calcium carbonate as balanced off the stoichiometry of the initial ion exchange.

Likewise, when determining the brucite content, the release of water linked to the transformation from brucite (MH) to magnesite was factored in as a fraction of the magnesite content ($\mathrm{LdC}_m$). The water loss associated with the hydrotalcite content was also subtracted from this calculation to prevent overestimation:

$$\mathrm{MH} = \left(\frac{58.32}{18.01}\right)\mathrm{Ldx_a} + \left(\frac{58.32}{44.01}\right)\mathrm{LdC}_m - \left(\frac{44.01}{531.92} + \frac{8 \times 18.01}{531.92}\right)\left(\frac{\mathrm{Ldh}_b}{0.12}\right) \tag{11}$$

**Table 7.** Weight of sample at different temperatures

| Mix | $W_{32^\circ}$ ($\mu$g) | $W_{105^\circ}$ ($\mu$g) | $W_{150^\circ}$ ($\mu$g) | $W_{230^\circ}$ ($\mu$g) | $W_{460^\circ}$ ($\mu$g) | $W_{530^\circ}$ ($\mu$g) | $W_{635^\circ}$ ($\mu$g) | $W_{1000^\circ}$ ($\mu$g) |
|---|---|---|---|---|---|---|---|---|
| SF10NH8 | 12,866.57 | 12,539.35 | 12,364.07 | 12,205.64 | 11,646.56 | 11,502.70 | 11,187.45 | 10,825.05 |
| SF20NH8 | 12,454.08 | 12,062.84 | 11,910.97 | 11,788.55 | 11,522.71 | 11,400.00 | 11,167.40 | 11,037.15 |
| SF10NH10 | 12,408.41 | 11,791.61 | 11,544.55 | 11,345.30 | 10,908.98 | 10,703.15 | 10,347.33 | 10,137.07 |
| SF20NH10 | 12,547.22 | 12,123.28 | 11,967.29 | 11,813.33 | 11,461.45 | 11,336.02 | 11,047.50 | 10,671.72 |

**Table 8.** Evaluation of bound water content, hydrotalcite, portlandite, and brucite phases

| Mix | Free water | Ldh (%) | | Ldx (%) | | LdC (%) | | | | | | | | |
|---|---|---|---|---|---|---|---|---|---|---|---|---|---|---|
| | $Ldh_a$ (%) | $Ldh_b$ | $Ldh_c$ | $Ldx_a$ (%) | $Ldx_b$ (%) | $LdC_m$ | $LdC_c$ | | | | | | | |
| | 32°C to 105°C | 105°C to 150°C | 150°C to 230°C | 230°C to 460°C | 460°C to 530°C | 530°C to 635°C | 635°C to 1,000°C | $LdC_a$[a] (%) | Bound water (%) | HT % | Free CH (%) | Free MH (%) | 28-day Strength (MPa) | 120-day Strength (MPa) |
| SF10NH8 | 2.543 | 1.398 | 1.264 | 4.459 | 1.147 | 2.514 | 2.890 | 1.898 | 8.48 | 11.65 | 6.29 | 13.65 | 29.76 | 31.37 |
| SF20NH8 | 3.141 | 1.259 | 1.015 | 2.204 | 1.017 | 1.928 | 1.080 | 1.853 | 4.75 | 10.49 | 2.96 | 5.98 | 30.74 | 31.57 |
| SF10NH10 | 4.971 | 2.095 | 1.690 | 3.700 | 1.746 | 3.018 | 1.783 | 2.038 | 9.16 | 17.46 | 6.78 | 9.80 | 35.10 | 41.33 |
| SF20NH10 | 3.379 | 1.287 | 1.270 | 2.903 | 1.035 | 2.380 | 3.100 | 1.993 | 6.72 | 10.72 | 5.98 | 8.76 | 32.82 | 39.17 |

[a]The calculation of $Ldc_a$ is shown in Table S2 of supplementary material.

where $(58.32/18.01)$ = molar mass ratio of $Mg(OH)_2$ to water; and $(58.32/44.01)$ = molar mass ratio of $Mg(OH)_2$ to $CO_2$. The percentage of water crystallization of hydrotalcite is 12%. Since $Ldh_b$ gives an estimation of the crystalline water lost from the hydrotalcite phase, $(Ldh_b/0.12)$ estimates the hydrotalcite content, and $(44.01/531.92 + 8 \times 18.01/531.92)$ is the molar mass ratio of $CO_2$ and water lost by Eq. (9).

Table 8 indicates that the bound water content related to the gel phases was most significant in the SF10NH10 sample, followed by SF10NH8, aligning with the findings from XRD analysis. Additionally, the hydrotalcite phases were more prominent in the NH10 samples, which is in accordance with the XRD results. While the values obtained through Rietveld refinement did not precisely coincide with the TG quantification values, the increasing order remained consistent for both methods. The portlandite content was higher in the case of SF10NH10, which might have had a favorable impact on the strength development, with 41.33 MPa at 120 days. While a higher portlandite content was observed in SF10NH8, it is worth noting that the same level of strength development was not achieved. This discrepancy could potentially be attributed to the higher amount of brucite. Both of these phases could contribute to the formation of silicate gel phases. However, it is essential to recognize that magnesium–silicate–hydrate (MSH) might not be as structurally robust as CSH.

Free water loss was indicated by the mass loss below 105°C, the first endotherm in the DTA. Unexpectedly, the mass loss for SF10NH10 was substantially higher, inconsistent with the strength findings. The water molecules might have been the end product of a chemical reaction over the course of 7 to 28 days. The XRD analysis revealed that the calcite peaks decreased as magnesite and dolomite phases formed at 28 days. Calcite is typically a stable compound. Calcite could, however, transform into magnesite and dolomites in the presence of magnesium (Walkley et al. 2017) if there was a localized high-temperature difference. The release of water molecules was typically associated with this phenomenon, explaining why more free water existed. The ascending order, SF10NH8 < SF20NH8 < SF20NH10 < SF10NH10 might thus correspond to the calcite's decreasing rate over 28 days.

## Conclusions

A pilot experimental investigation was carried out to examine how an alkali-activated blend of silica fume and slag evolved in terms of flow and strength. The primary objective was to assess the potential of utilizing less pure industrial-grade soda ash and hydrated lime to activate this blend efficiently. Furthermore, two alternative mixing methods were employed and evaluated for their efficacy. Additional microstructural and thermal analyses of the one-part mixes were conducted to gain a deeper understanding of the strength development of the proposed mixture. Directly mixing industrial-grade soda ash and hydrated lime with precursors offered a practical and cost-effective substitute for analytical-grade activators. Further research is needed to design concrete mixes and investigate performance-based properties using this binder.

The key findings of the study are as follows:

1. Industrial-grade SA and HL present a substantial cost advantage compared to analytical $Na_2CO_3$ and $Ca(OH)_2$, with potential cost savings of up to 94.5%. Moreover, when industrial-grade SA-HL is compared to an analogous NaOH solution-activated mixture, a cost reduction of 91% is observed.
2. The increase in SA-HL content initially led to a reduction in both yield stress and plastic viscosity. However, when additions exceeded 10%, the yield stress increased, indicating a complex relationship between NaOH content and flow properties. Including silica fume (SF) beyond 10% in the mix increased the yield stress and decreased flowability.
3. The soda ash's role differed depending on whether it was added as a premixed solution or with hydrated lime. Soda ash acted as an accelerator in premixed solutions, causing quick or false-setting. The plastic viscosity and yield stress were notably high, hence a less fluid mixture. The readily available $CO_3^{2-}$ ions in solution form subdued the dissolution process in the long run. Although premixing led to increased initial strength, the formation of carbonates created barriers to subsequent reactions impacting later-age strength development. However, higher $Na^+$ content was advantageous for strength development.
4. A different mechanism occurred when soda ash was added to hydrated lime. Initially, soda ash reacted with hydrated lime to release $Na^+$, $OH^-$, $Ca^{2+}$, and $CO_3^{2-}$ ions. These ions interacted differently, with $Ca^{2+}$ ions from hydrated lime exhausting the $CO_3^{2-}$ ions, potentially allowing $Na^+$ ions to react with slag particles, lowering the yield stress, promoting stable polymerization reactions, and contributing to overall strength development.
5. Analytical-grade chemical activated mixes demonstrated higher strength due to enhanced reactivity of the activators from finer particle size. However, combining these finer reagents with more than 10% SF would increase water demand as a whole, resulting in insufficient wetting of slag particles and a subsequent decrease in strength.
6. In the XRD analysis of the one-part mortars activated with industrial-grade SA and HL, the coexistence of CSH, CASH, and NASH shaped the 28-day strength. Phases of calcite and hydrotalcite also contributed to the development of strength. The SEM and TG/DTA further validated the results.
7. Reduced SF content favored the development of a microstructure dominated by CSH and NASH, whereas higher SF content led to the coexistence of CSH, CASH, and NASH.
8. The blends exhibited the most robust strength improvement and impressive rheological properties and flow when incorporating 10% SF and targeting a 10% NaOH content with a water-to-solid (w/s) ratio of 0.45. This mix achieved a strength of 35.1 MPa after 28 days and 41.33 MPa after 120 days, establishing its suitability for structural applications.

## Data Availability Statement

All data, models, and code generated or used during the study appear in the published article.

## Supplemental Materials

Figs. S1–S8 and Tables S1–S2 are available online in the ASCE Library (www.ascelibrary.org).